\documentclass[prb,twocolumn,showpacs,preprintnumbers,amsmath,aps]{revtex4}
\usepackage{graphicx,hyperref}
\usepackage{bm}
\usepackage{subfigure}

\newcommand\etab{\bar \eta}

\def\nbC{{\mathchoice {\setbox0=\hbox{$\displaystyle\rm C$}%
\hbox{\hbox to0pt{\kern0.4\wd0\vrule height0.9\ht0\hss}\box0}}
{\setbox0=\hbox{$\textstyle\rm C$}\hbox{\hbox
to0pt{\kern0.4\wd0\vrule height0.9\ht0\hss}\box0}}
{\setbox0=\hbox{$\scriptstyle\rm C$}\hbox{\hbox
to0pt{\kern0.4\wd0\vrule height0.9\ht0\hss}\box0}}
{\setbox0=\hbox{$\scriptscriptstyle\rm C$}\hbox{\hbox
to0pt{\kern0.4\wd0\vrule height0.9\ht0\hss}\box0}}}}
\def\nbQ{{\mathchoice {\setbox0=\hbox{$\displaystyle\rm
Q$}\hbox{\raise
0.15\ht0\hbox to0pt{\kern0.4\wd0\vrule height0.8\ht0\hss}\box0}}
{\setbox0=\hbox{$\textstyle\rm Q$}\hbox{\raise
0.15\ht0\hbox to0pt{\kern0.4\wd0\vrule height0.8\ht0\hss}\box0}}
{\setbox0=\hbox{$\scriptstyle\rm Q$}\hbox{\raise
0.15\ht0\hbox to0pt{\kern0.4\wd0\vrule height0.7\ht0\hss}\box0}}
{\setbox0=\hbox{$\scriptscriptstyle\rm Q$}\hbox{\raise
0.15\ht0\hbox to0pt{\kern0.4\wd0\vrule height0.7\ht0\hss}\box0}}}}
\def\nbT{{\mathchoice {\setbox0=\hbox{$\displaystyle\rm
T$}\hbox{\hbox to0pt{\kern0.3\wd0\vrule height0.9\ht0\hss}\box0}}
{\setbox0=\hbox{$\textstyle\rm T$}\hbox{\hbox
to0pt{\kern0.3\wd0\vrule height0.9\ht0\hss}\box0}}
{\setbox0=\hbox{$\scriptstyle\rm T$}\hbox{\hbox
to0pt{\kern0.3\wd0\vrule height0.9\ht0\hss}\box0}}
{\setbox0=\hbox{$\scriptscriptstyle\rm T$}\hbox{\hbox
to0pt{\kern0.3\wd0\vrule height0.9\ht0\hss}\box0}}}}
\def\nbS{{\mathchoice
{\setbox0=\hbox{$\displaystyle     \rm S$}\hbox{\raise0.5\ht0%
\hbox to0pt{\kern0.35\wd0\vrule height0.45\ht0\hss}\hbox
to0pt{\kern0.55\wd0\vrule height0.5\ht0\hss}\box0}}
{\setbox0=\hbox{$\textstyle        \rm S$}\hbox{\raise0.5\ht0%
\hbox to0pt{\kern0.35\wd0\vrule height0.45\ht0\hss}\hbox
to0pt{\kern0.55\wd0\vrule height0.5\ht0\hss}\box0}}
{\setbox0=\hbox{$\scriptstyle      \rm S$}\hbox{\raise0.5\ht0%
\hboxto0pt{\kern0.35\wd0\vrule height0.45\ht0\hss}\raise0.05\ht0%
\hbox to0pt{\kern0.5\wd0\vrule height0.45\ht0\hss}\box0}}
{\setbox0=\hbox{$\scriptscriptstyle\rm S$}\hbox{\raise0.5\ht0%
\hboxto0pt{\kern0.4\wd0\vrule height0.45\ht0\hss}\raise0.05\ht0%
\hbox to0pt{\kern0.55\wd0\vrule height0.45\ht0\hss}\box0}}}}
\def\nbZ{{\mathchoice {\hbox{$\sf\textstyle Z\kern-0.4em Z$}}
{\hbox{$\sf\textstyle Z\kern-0.4em Z$}}
{\hbox{$\sf\scriptstyle Z\kern-0.3em Z$}}
{\hbox{$\sf\scriptscriptstyle Z\kern-0.2em Z$}}}}

\begin{document}

\title{Nonperturbative Functional Renormalization Group for Random Field Models. IV: Supersymmetry and its spontaneous breaking.}

\author{Matthieu Tissier} \email{tissier@lptl.jussieu.fr}
\affiliation{LPTMC, CNRS-UMR 7600, Universit\'e Pierre et Marie Curie,
bo\^ite 121, 4 Place Jussieu, 75252 Paris c\'edex 05, France}

\author{Gilles Tarjus} \email{tarjus@lptl.jussieu.fr}
\affiliation{LPTMC, CNRS-UMR 7600, Universit\'e Pierre et Marie Curie,
bo\^ite 121, 4 Place Jussieu, 75252 Paris c\'edex 05, France}

\date{\today}

\begin{abstract}
We apply the nonperturbative functional renormalization group (NP-FRG) in the superfield formalism that we have developed in the preceding paper to study long-standing issues concerning the critical behavior of the random field Ising model. Through the introduction of an appropriate regulator and a supersymmetry-compatible nonperturbative approximation, we are able to follow the supersymmetry, more specifically the superrotational invariance first unveiled by Parisi and Sourlas [Phys. Rev. Lett. \textbf{43}, 744 (1979)], and its spontaneous breaking along the RG flow. Breaking occurs below a critical dimension $d_{DR}\simeq 5.1$, and the supersymmetry-broken fixed point that controls the critical behavior then leads to a breakdown of the ``dimensional reduction'' property. We solve the NP-FRG flow equations numerically and determine the critical exponents as a function of dimension down to $d\lesssim3$, with a good agreement in $d=3$ and $d=4$ with the existing numerical estimates.
\end{abstract}


\maketitle

\section{Introduction}
\label{sec:introduction}

Quenched disorder induces an extrinsic inhomogeneity in otherwise translationally invariant pure systems. As a result, the equilibrium and out-of-equilibrium physics of disordered systems may be influenced by rare collective events, such as ``avalanches'', statistically unlikely regions, such as ``droplets'' or ``Griffiths regions'', and by the proliferation of ``metastable states''. Our objective is to develop a theory describing the long-distance physics of such systems through a nonperturbative functional renormalization group (NP-FRG) method and our first focus is the equilibrium behavior of the random-field model. We  showed in previous papers, which we refer to as I\cite{tarjus08} and II\cite{tissier08}, that the effect of avalanches and droplets can be captured by an approach based on the RG flow of the cumulants of the renormalized disorder, provided that the full functional dependence of the latter is accounted for.\cite{footnote00} A functional RG is therefore required to let the singular behavior due to rare events or regions emerge along the flow, as also shown in the case of the equilibrium and forced behavior of manifolds in a random environment.\cite{fisher86a,balents-fisher93,balents96,BLchauve00,CUSPledoussal,BLbalents04,CUSPledoussal09,BLledoussal10}

The issue of metastable states, which, at zero temperature where the concept is well defined, refers to the presence of many minima of the microscopic hamiltonian (bare action) not simply related by symmetry transformations, is a recurring conundrum in theories of disordered systems. In the case of the random-field Ising model (RFIM) under study, it has a clear manifestation. The critical behavior of the model being controlled by a zero-temperature fixed point,\cite{villain84,fisher86,nattermann98} the long-distance physics can be described through the properties of the ground state which, due to the presence of the random field, is obtained as the solution of a stochastic field equation. By standard field-theoretic manipulations, this leads to a theory expressed in terms of superfields. Parisi and Sourlas\cite{parisi79} showed that a supersymmetry of the theory, more specifically the invariance under rotations of the underlying superspace, implies the property of dimensional reduction, according to which the critical behavior of the RFIM in $d$ dimensions is identical to that of the pure Ising model in $d-2$ dimensions. The property however has been proven to be wrong in low enough dimension.\cite{imbrie84,bricmont87} At the same time, it has been understood that the superfield construction breaks down because of the presence of many solutions of the stochastic field equation.\cite{parisi84b} However, in the Parisi-Sourlas formalism, it is not possible to disentangle breaking of superrotational invariance and collapse of the formalism due to the appearance of multiple solutions.

In the companion paper, henceforth referred to as paper III,\cite{tissier11b} we have shown how to resolve the above conundrum and to combine an extended superfield approach with the NP-FRG formalism. In particular, ground-state selection can be achieved by adding a weighting factor involving an auxiliary temperature and letting at the end of the manipulations the auxiliary temperature go to zero in the exact NP-FRG equations for the cumulants of the renormalized disorder. The resulting property of ``Grassmannian ultralocality'' then allows one to  specifically investigate supersymmetry (superrotational invariance) and its spontaneous breaking along the RG flow. Such an investigation is the purpose of the present paper.

The outline of the article is as follows. In Sec.~\ref{sec:model}, we briefly summarize the main steps of the NP-FRG in a superfield formalism that we have developed in paper III. This allows us to recall definitions, notations, and RG equations that will be used in the present article. In particular, we stress two important, and distinct, formal properties of  the superfield theory: ``Grassmannian ultralocality" and ``superrotational invariance". We also consider the RG flow of the Ward-Takahashi identities associated with the latter and the consequences for the choice of the infrared regulator. In Sec.~\ref{sec:SUROT invariance}, we show through our NP-FRG formalism that the superrotational invariance nonperturbatively leads to dimensional reduction. We conclude the section by building a scenario for a spontaneous breaking of the superrotational invariance along the RG flow that is based on our previous results on the appearance of a ``cusp" in the functional dependence of the cumulants of the renormalized random field,\cite{tarjus08,tissier08,tarjus04,tissier06} and we propose a continuation of the NP-FRG flow equations when spontaneous breaking has taken place. 

Sec.~\ref{sec:approximation scheme} is devoted to the development of a supersymmetry-compatible nonperturbative approximation scheme for the exact NP-FRG equations. It relies on combined truncations of the cumulant expansion and the expansion in spatial derivatives of the field. We also provide details on the numerical resolution. In Sec.~\ref{sec:results} we present the results. We show in particular that breakdown of the dimensional reduction predictions for the critical exponents of the RFIM occurs below a critical dimension $d_{DR}\simeq 5.1$. We compute the critical exponents as a function of dimension down to $d=3$ and we find good agreement with the best available estimates in $d=3$ and $d=4$. Finally, we show that scaling is described by three  independent exponents, contrary to a proposed conjecture.\cite{schwartz85b,soffer85,schwartz91}

A short account of this work has appeared in Ref.[\onlinecite{tissier11}].

\section{NP-FRG in the superfield formalism for the RFIM}
\label{sec:model}

\subsection{Summary}

We start by briefly recalling the main features of the formalism presented in the preceding article, with the associated definitions and notations.

In paper III,\cite{tissier11b} we have developed a NP-FRG theory for describing the equilibrium long-distance physics of the RFIM that is based on an extension of the Parisi-Sourlas\cite{parisi79} supersymmetric formalism. The latter relies on the fact that the critical behavior of the model is dominated by disorder-induced fluctuations, thermal fluctuations being subdominant, and can therefore be studied by looking directly at zero temperature. The equilibrium properties are then described by the ground state which is solution of the following stochastic field equation
\begin{equation}
\label{eq_stochastic}
\dfrac{\delta S[\varphi;h]}{\delta \varphi(x)} = J(x),
\end{equation}
where we have added an external source (a magnetic field) $J$ conjugate to the $\varphi$ field and the action $S[\varphi;h]$ is given by
\begin{equation}
\label{eq_ham_dis}
S= \int_{x} \bigg\{ \frac
{1}{2}  \left(\partial_{\mu} \varphi(x) \right) ^2 + U_B(\varphi(x)) -
h(x) \varphi(x) \bigg\} ,
\end{equation}
where $ \int_{x} \equiv \int d^d x$,  $U_B(\varphi)= (\tau/2) \varphi^2  +  (u/4!) \varphi^4$, 
with $h(x)$ a random  (magnetic) field sampled from a Gaussian distribution of zero mean and variance $\overline{h(x)h(y)}=\Delta_B \ \delta^{(d)}(x-y)$.

Two key ingredients of our extension of the Parisi-Sourlas formalism are:

(1) the need to consider multiple copies (or replicas) of the original
system with the same disorder, each copy being coupled to a different
applied source, in order to generate cumulants of the renormalized
disorder with their full functional dependence, thereby allowing for
the emergence of a nonanalytic behavior in the field arguments,

(2) the introduction of a weighting factor involving an auxiliary
temperature $\beta^{-1}$ to the solutions of the stochastic field
equation, so that when $\beta^{-1}$ appraches 0 only the ground state contributes to the generating functional.

By using standard field-theoretical techniques,\cite{zinnjustin89} one then ends up with a superfield theory for multiple copies in a curved superspace, with
\begin{equation}
\begin{aligned}
  \label{eq_part_func_multicopy}
&\mathcal Z^{(\beta)}[\{\mathcal J_a\}]=\exp (\mathcal W^{(\beta)}[\{\mathcal J_a\}])=\overline{\exp(\sum_{a=1}^n\mathcal W_h^{(\beta)}[ \mathcal J_a])} \\&= \int \prod_{a=1}^{n}\mathcal Q\Phi_a  \exp \bigg(-S^{(\beta)}[\{\Phi_a\}] +  \sum_{a=1}^{n} \int_{\underline x}  \mathcal J_a(\underline x)  \Phi_a(\underline x)\bigg),
\end{aligned}
\end{equation}
where the multicopy action is given by
\begin{equation}
\begin{aligned}
\label{eq_superaction_multicopy}
S^{(\beta)}&[\{\Phi_a\}] = \sum_{a=1}^{n} \int_{\underline{x}} \left[ \frac 12 (\partial_{\mu}\Phi_a(\underline{x}))^2+U_{B}(\Phi_a(\underline{x}))\right] \\&- \frac{\Delta_B}{2}\sum_{a_1=1}^{n}\sum_{a_2=1}^{n} \int_{x}\int_{\underline{\theta}_1\underline{\theta}_2} \Phi_{a_1}(x,\underline{\theta}_1)\Phi_{a_2}(x,\underline{\theta}_2).
\end{aligned}
\end{equation}
In the above equations, we have introduced a superspace (coordinates $\underline x$) comprising the $d$-dimensional Euclidean space (coordinates $x$) and a $2$-dimensional Grassmannian space (anticommuting coordinates $\underline \theta =\{\theta, \bar \theta \}$); the metric is flat in the Euclidean sector and curved in the Grassmannian one with the curvature proportional to $\beta$. 
(For instance, the integral over superspace is defined as $\int_{\underline x}\equiv\int_{x}\int_{\underline{\theta}} \equiv \int d^dx \int \int (1+\beta \bar\theta \theta)d\theta d\bar\theta$.) We have also defined superfields $\Phi_a(\underline x)=\varphi_a(x) +\bar\theta \psi_a(x) +\bar\psi_a(x) \theta + \bar \theta \theta \hat\phi_a(x)$, with one auxiliary bosonic (``response") field $\hat \phi_a$ and two auxiliary fermionic (``ghost") fields $\bar\psi_a$ and $\psi_a$, and associated supersources $\mathcal J_a(\underline x)=J_a(x) +\bar \theta K_a(x)+\bar K_a(x) \theta+\bar \theta \theta \hat J_a(x)$. The action in Eq. (\ref{eq_superaction_multicopy}) is invariant under a large group of (bosonic and fermionic) symmetries.

When expanded in increasing number of sums over copies,  the generating functional of the connected Green's functions,  $\mathcal W^{(\beta)}[\{\mathcal J_a\}]$, gives access to the cumulants of the random generating functional $\mathcal W_h^{(\beta)}[\mathcal J]$. 

We have next applied the NP-FRG formalism to this superfield theory. This proceeds by first adding an infrared (IR) regulator that enforces a progressive account of the fluctuations to the bare action,
\begin{equation}
\begin{aligned}
\label{eq_regulator}
\Delta S_k^{(\beta)}=&\frac 12  \sum_{a}\int_{x_1 x_2}\int_{\underline \theta}
\Phi_{a}(x_1,\underline \theta)\widehat{R}_k(|x_1-x_2|)\Phi_{a}(x_2,\underline{\theta})\\&+\frac 12 \sum_{a_1,a_2}\int_{\underline{x}_1 \underline{x}_2} \Phi_{a_1}(\underline{x}_1)\widetilde{R}_k(|x_1-x_2|) \Phi_{a_2}(\underline{x}_2),
\end{aligned}
\end{equation}
where the two cutoff functions $\widehat{R}_k$ and $\widetilde{R}_k$ are related through
\begin{equation}
\label{eq_cutoffs_relation}
\widetilde{R}_k(q^2)=-\frac{\Delta_k}{Z_k}\partial_{q^2}\widehat{R}_k(q^2),
\end{equation}
with $q$ the Euclidean momentum, $\Delta_k$ the strength of the renormalized random field, and $Z_k$ the field renormalization constant. This ensures that all symmetries of the theory  are satisfied. This includes the superrotational invariance found when the theory is restricted to a single copy and to $\beta=0$: it then corresponds to $\Delta_k=\Delta_B Z_k$, a property that is valid at the microscopic (UV) scale $\Lambda$ (see also below).

One next introduces the effective average action\cite{wetterich93,berges02} $\Gamma_k^{(\beta)}[\{\Phi_a\}]=-\mathcal W_k^{(\beta)}[\{\mathcal J_a\}]+\sum_a\int_{\underline x} \Phi_a(\underline x) \mathcal J_a(\underline x) -\Delta S_k^{(\beta)}[\{\Phi_a\}]$, whose dependence on the IR cutoff $k$ is governed by an exact renormalization-group equation (ERGE).
The effective average action can also be expanded in increasing number of sums over copies, each term of the expansion being then related through the Legendre transform to the cumulants of $\mathcal W_h^{(\beta)}[\mathcal J]$. We refer to the $p$th order term of the expansion of $\Gamma_k^{(\beta)}[\{\Phi_a\}]$ as the $p$th  ``cumulant of the renormalized disorder''. These expansions in increasing number of sums over copies lead to systematic algebraic manipulations that allow one to derive from the ERGE for $\Gamma_k^{(\beta)}[\{\Phi_a\}]$ a hierarchy of coupled ERGE's for the cumulants of the renormalized disorder.

We have unveiled an important property of the random generating functional $\mathcal W_h^{(\beta)}[\mathcal J]$: the latter satisfies a specific dependence on the Grassmann coordinates, which we have called ``Grassmannian ultralocality", if and only if a unique solution of the stochastic field equation is included in its computation. This translates into an ``ultralocal" property of the cumulants of the renormalized disorder. ``Grassmannian ultralocality" becomes a property of the superfield theory when  $\beta \rightarrow \infty$; it is also asymptotically found for finite $\beta$ when $k\rightarrow 0$ (after going to dimensionless quantities). It then reflects the desired ground-state dominance. When this property is satisfied (\textit{e.g.} by setting $\beta^{-1}=0$), the ERGE's for the cumulants simplify and only involve ''ultralocal" quantities that can be evaluated for physical fields $\Phi_a(\underline x)=\phi_a(x)$, \textit{i.e.} for superfields that are uniform in the Grassmann subspace.

For illustration, we give below the ERGE's for the first two cumulants under the property of ``Grassmannian ultralocality", which will be needed in the following:
\begin{equation}
\label{eq_flow_Gamma1_ULapp}
\begin{split}
&\partial_t\Gamma_{k1}\left[\phi_1\right ]= \\&-
\dfrac{1}{2} \tilde{\partial}_t \int_{x_2x_3}\widehat{P}_{k;x_2x_3}[\phi_1] \big(\Gamma_{k2;x_2,x_3}^{(11)}\left[\phi_1,\phi_1\right ] - \widetilde{R}_{k;x_2x_3}\big)
\end{split}
\end{equation}
and
\begin{equation}
\label{eq_flow_Gamma2_ULapp}
\begin{split}
&\partial_t\Gamma_{k2}\left[\phi_1,\phi_2\right ]=\\&
\dfrac{1}{2} \tilde{\partial}_t \int_{x_3x_4}\big \{- \widehat{P}_{k;x_3x_4}\left[\phi_1\right ] \Gamma_{k3;x_3,.,x_4}^{(101)}\left[\phi_1,\phi_2,\phi_1\right ]+\\& \widetilde{P}_{k;x_3x_4}\left[\phi_1,\phi_1\right ] \Gamma_{k2;x_3x_4,.}^{(20)}\left[\phi_1,\phi_2\right ]+ \frac{1}{2}\widetilde{P}_{k;x_3x_4}\left[\phi_1,\phi_2\right ] \\& \times \left( \Gamma_{k2;x_3,x_4}^{(11)}\left[\phi_1,\phi_2\right ] - \widetilde{R}_{k;x_3x_4}\right) + perm(12)\big \},
\end{split}
\end{equation}
where $perm (12)$ denotes the expression obtained by permuting $\phi_1$ and $\phi_2$, $t=\log(k/\Lambda)$,  $\widetilde{\partial}_t$ is a short-hand notation to indicate a derivative acting only on the cutoff functions (\textit{i.e.},  $\widetilde{\partial}_t \equiv \partial_t \widehat{R}_k\, \delta/\delta \widehat{R}_k + \partial_t \widetilde{R}_k \, \delta/\delta \widetilde{R}_k$), and the propagators $\widehat{P}_{k}$ and $\widetilde{P}_{k}$ (denoted by $\widehat{P}_{k}^{[0]}$ and  $\widetilde{P}_{k}^{[0]}$ in  paper III\cite{tissier11b}) are defined as
\begin{equation}
\label{eq_hatP_zero}
\widehat {P}_{k}[\phi ]=\left(\Gamma _{k,1}^{(2)}[ \phi ]+\widehat R_k\right) ^{-1}
\end{equation}
and
\begin{equation}
\label{eq_tildeP_zero}
\widetilde {P}_{k}[\phi_1, \phi_2 ]= \widehat {P}_{k}[ \phi_1 ](\Gamma _{k,2}^{(11)}[\phi_1, \phi_2 ]-\widetilde R_k ) \widehat {P}_{k}[ \phi_2 ],
\end{equation}
and superscripts indicate functional differentiation with respect to the field arguments.

Generically, the flow of $\Gamma_{kp}\left[\phi_1,...,\phi_p\right ]$
involves three types of quantities: the propagators  $\widehat{P}_{k}$
and  $\widetilde{P}_{k}$, second functional derivatives of
$\Gamma_{kp}$ in which all the arguments are different, and second
functional derivatives of $\Gamma_{k(p+1)}$ with two of their
arguments equal to each other. A graphical representation of the
hierarchy of ERGE's is provided is Appendix C of the companion paper III. 

In the present paper we will focus on these flow equations, which correspond to the $\beta \rightarrow \infty$ limit and describe the equilibrium properties associated with the ground state (recall that the ``bath'' temperature is also at $T=0$). We will briefly comment on the RG flow of corrections to ``Grassmannian ultralocality" when $\beta$ is large but finite. 

Finally, it is worth stressing that to obtain the flow equation for $\Gamma_{kp}[\phi_1,...,\phi_p]$ with its full functional dependence on the $p$ field arguments, one needs to consider at least $p$ copies. If one considers less than $p$ copies, the equation necessarily involves $\Gamma_{kp}$ and its derivatives in which several of the arguments are equal. As we will show later, breaking of supersymmetry and breakdown of dimensional reduction are precisely related to the analyticity properties of the functionals when taking the limit of equal arguments. Formally, the whole hierarchy of flow equations for the cumulants can thus be obtained by considering an arbitrary large number of copies.

\subsection{Superrotational invariance, Ward-Takahashi identities and RG flow}

When ``Grassmannian ultralocality" is satisfied, the curvature $\beta$ disappears from the flow equations [see Eqs.~(\ref{eq_flow_Gamma1_ULapp},\ref{eq_flow_Gamma2_ULapp})]. Formally, the latter are then the same as those obtained from considering a flat Grassmann subspace. We have shown in paper III that in this case, and provided one restricts the supersources such that one effectively recovers a one-copy theory, the theory is invariant under ``superrotations" that mix the Euclidean and Grassmannian sectors of the superspace. The associated generators are $\overline{ \mathcal Q}_{\mu}=-x_{\mu} \partial_{\theta}+(2/\Delta_B) \bar \theta \partial_{\mu}$ and $\mathcal Q_{\mu}= x_{\mu} \partial_{\bar\theta}+(2/\Delta_B)\theta \partial_{\mu}$. As any linearly realized continuous symmetry, superrotational invariance leads to a set of Ward-Takahashi (WT) identities for the one-particle irreducible (1PI) generating functional, \textit{i.e.} at scale $k$ the effective average action. For a flat superspace and a restriction to one copy, the WT identity for the effective average action reads
\begin{equation}
\label{eq_ward_k}
\int_{\underline{x}}\Phi(\underline{x})\mathcal Q_{\mu}\Gamma_{k;\underline{x}}^{(1)}[\Phi]=0
\end{equation}
and similarly with $\overline{\mathcal Q}_\mu$.

One can now check that the above WT identities are \textit{a priori} stable under the RG flow, \textit{i.e.}
 \begin{equation}
\label{eq_flow_ward}
\partial_t\int_{\underline{x}}\Phi(\underline{x})\mathcal Q_{\underline{x}}\Gamma_{k;\underline{x}}^{(1)}[\Phi]=0,
 \end{equation}
where $\mathcal Q_{\underline{x}}$ generically indicates a component
$\mathcal Q_{\mu}$ and, since there is no curvature, $\int_{\underline x}\equiv \int d^dx \int d\theta d\bar \theta$. To prove Eq.~(\ref{eq_flow_ward}), we rewrite the ERGE for the effective average action when the superfield theory is restricted to one copy and is considered in a flat superspace ($\beta=0$):
\begin{equation}
\label{eq_ERGE_functional}
\partial_t \Gamma_k[\left\lbrace \Phi_a \right\rbrace]=\frac 12 \int_{\underline{x}_1\,\underline{x}_2}   \left( \partial_t \mathcal R_{k;\underline x_1,\underline x_2}\right)  \mathcal P_{k;\underline x_1,\underline x_2}[\left\lbrace \Phi_a \right\rbrace] ,
\end{equation}
where the regulator function $\mathcal R_{k;\underline x_1,\underline x_2}=\delta_{\bar \theta_1,\bar\theta_2}\widehat R_k(\vert x_1-x_2\vert)+\widetilde R_k(\vert x_1-x_2\vert)$ with $\delta_{\bar \theta_1,\bar\theta_2}=(\bar \theta_1-\bar \theta_2)(\theta_1- \theta_2)$ and the modified (full) propagator $\mathcal P_k$ is the inverse in the sense of operators of $\Gamma_k^{(2)}+\mathcal R_k$. (Inserting the ``multilocal'' expansion in the Grassmann coordinates and using the assumption of  ``Grassmannian ultralocality" then leads to ERGE's for the cumulants which are just Eqs.~(\ref{eq_flow_Gamma1_ULapp},\ref{eq_flow_Gamma2_ULapp}) and their extension to higher orders with all copy superfields taken as equal in both sides of the equations.) 

The flow of the left-hand side of Eq.~(\ref{eq_flow_ward}) can then be expressed as
 \begin{equation}
\begin{aligned}
\int_{\underline{x}}\Phi(\underline{x})&\mathcal
Q_{\underline{x}}\partial_t \Gamma_{k;\underline{x}}^{(1)}[\Phi]\\
&= \frac{1}{2}\tilde{\partial}_t
\int_{\underline{x}_1\,\underline{x}_2\,
  \underline{x}_3}\Phi(\underline{x}_1)\mathcal Q_{\underline{x}_1}
\left( \mathcal
P_{k;\underline{x}_2\underline{x}_3}[\Phi]\Gamma_{k;\underline{x}_3\underline{x}_2\underline{x}_1}^{(3)}[\Phi]\right)
\\
& =\frac{1}{2}\tilde{\partial}_t \int_{\underline{x}_2\,\underline{x}_3}\mathcal P_{k;\underline{x}_2\underline{x}_3}[\Phi] \int_{\underline{x}_1}\Phi(\underline{x}_1)\mathcal Q_{\underline{x}_1} \Gamma_{k;\underline{x}_3\underline{x}_2\underline{x}_1}^{(3)}[\Phi],
\end{aligned}
\end{equation}
where we have used that $\mathcal Q_{\underline{x}_1}$ does not act on $\mathcal P_{k;\underline{x}_2\underline{x}_3}$. Assuming that the WT identity, Eq.~(\ref{eq_ward_k}), is satisfied down to the scale $k$ (we study its further evolution), one has
 \begin{equation}
\begin{aligned}
\int_{\underline{x}_1}\Phi(\underline{x}_1)\mathcal Q_{\underline{x}_1} \Gamma_{k;\underline{x}_3\underline{x}_2\underline{x}}^{(3)}[\Phi]= - \left( \mathcal Q_{\underline{x}_2} + \mathcal Q_{\underline{x}_3}\right) \Gamma_{k;\underline{x}_3\underline{x}_2}^{(2)}[\Phi],
\end{aligned}
\end{equation}
which is obtained by  differentiating twice Eq.~(\ref{eq_ward_k}). After an integration by parts and a relabel of the dummy variables $\underline{x}_2,\underline{x}_3$, one then finds
 \begin{equation}
\begin{aligned}
\label{eq_dem_SUSY_DR}
\int_{\underline{x}}&\Phi(\underline{x})\mathcal Q_{\underline{x}}\partial_t \Gamma_{k;\underline{x}}^{(1)}[\Phi]\\&= -\frac{1}{2}\tilde{\partial}_t \int_{\underline{x}_2\,\underline{x}_3}\Gamma_{k;\underline{x}_3\underline{x}_2}^{(2)}[\Phi]\; \left(\mathcal Q_{\underline{x}_2} + \mathcal Q_{\underline{x}_3}\right) \mathcal P_{k;\underline{x}_2\underline{x}_3}[\Phi].
\end{aligned}
\end{equation}
Provided that one chooses the infrared cutoff function such that
 \begin{equation}
\label{eq_condition_cutoff}
\left( \mathcal Q_{\underline{x}_1} + \mathcal Q_{\underline{x}_2}\right) \mathcal R_{k;\underline{x}_1\underline{x}_2}=0,
\end{equation}
one can replace
$\Gamma_{k;\underline{x}_3\underline{x}_2}^{(2)}[\Phi]$ by
$\Gamma_{k;\underline{x}_3\underline{x}_2}^{(2)}[\Phi]+ \mathcal
R_{k;\underline{x}_3\underline{x}_2}$ in
Eq.~(\ref{eq_dem_SUSY_DR}). By using the fact that the latter is equal
to the inverse modified propagator $(\mathcal
P_{k}^{-1}[\Phi])_{\underline{x}_3\underline{x}_2}$, one can rewrite the right-hand side of Eq.~(\ref{eq_dem_SUSY_DR}) (up to the trivial factor of $-1/2$) as
\begin{equation}
\begin{aligned}
&\tilde{\partial}_t \int_{\underline{x}_2\,\underline{x}_3} (\mathcal P_{k}^{-1}[\Phi])_{\underline{x}_3\underline{x}_2}\; \mathcal Q_{\underline{x}_2} \mathcal P_{k;\underline{x}_2\underline{x}_3}[\Phi]\\&=\tilde{\partial}_t \int_{\underline{x}_2\,\underline{x}_4}\delta_{\underline{x}_2\underline{x}_4}\int_{\underline{x}_3}(\mathcal P_{k}^{-1}[\Phi])_{\underline{x}_3\underline{x}_4}\; \mathcal Q_{\underline{x}_2} \mathcal P_{k;\underline{x}_2\underline{x}_3}[\Phi]\\&=\tilde{\partial}_t \int_{\underline{x}_2\,\underline{x}_4}\delta_{\underline{x}_2\underline{x}_4}
\mathcal Q_{\underline{x}_2}\; \int_{\underline{x}_3}(\mathcal P_{k}^{-1}[\Phi])_{\underline{x}_3\underline{x}_4} \mathcal P_{k;\underline{x}_2\underline{x}_3}[\Phi]\\&=\tilde{\partial}_t \int_{\underline{x}_2\,\underline{x}_4}\delta_{\underline{x}_2\underline{x}_4}
\mathcal Q_{\underline{x}_2}\; \delta_{\underline{x}_2\underline{x}_4}.
\end{aligned}
\end{equation}
This last expression is easily shown to be identically zero, so that one finally obtains that Eq.~(\ref{eq_flow_ward}) is satisfied at scale $k$. The same reasoning can be repeated with the identity associated with $\overline{\mathcal Q}_{\underline{x}}$.

Therefore, if one starts with an initial condition that is superrotationally invariant, which can be enforced by suppressing fluctuations (see section II-A and paper III), the above derivation proves that the WT identities associated with the superrotational invariance are \textit{a priori} preserved along the RG flow if the regulator satisfies Eq.~(\ref{eq_condition_cutoff}). The latter condition can be reexpressed as
\begin{equation}
\begin{aligned}
&\left (\mathcal Q_{1 \mu}+ \mathcal Q_{2\mu} \right ) \mathcal R_{k;\underline x_1\underline x_2}=\left (\theta_1-\theta_2 \right )\bigg[ (x_{1\mu}-x_{2\mu}) \widehat R_k(\vert x_1-x_2\vert)\\&+\frac{2}{\Delta_B} \partial_{1\mu}\widetilde R_k(\vert x_1-x_2\vert)\bigg]=0,
\end{aligned}
\end{equation}
where we have used the fact that $\widehat R_k$ and $\widetilde R_k$ are translationally invariant functions of the Euclidean coordinates, with therefore $\partial_{1\mu}\widehat R_k=- \partial_{2\mu}\widehat R_k$ and similarly for $\widetilde R_k$. Going to Fourier space, this implies that $\widehat R_k$ and $\widetilde R_k$ are related through
\begin{equation}
\label{eq_relationSUSY_cutoffs}
\widetilde R_k(q^2)=-\Delta_B\, \partial_{q^2}\widehat R_k(q^2),
\end{equation}
which, as already stated, corresponds to Eq.~(\ref{eq_cutoffs_relation}) with $\Delta_k=\Delta_B Z_k$, a condition that we choose to enforce at the beginning of the flow when $k=\Lambda$. Such a choice of cutoff then avoids an \textit{explicit} breaking of  superrotational invariance.

\section{Superrotational invariance and its spontaneous breaking}
\label{sec:SUROT invariance}

\subsection{SUSY and dimensional reduction}

An important property of the theory is that the superrotational invariance (which for simplicity will often be denoted by the acronym SUSY, for supersymmetry, in the following) leads to dimensional reduction. We consider here the case where ``Grassmannian ultralocality" is satisfied so that the curvature $\beta$ drops out of the flow equations as in Eqs.~(\ref{eq_flow_Gamma1_ULapp},\ref{eq_flow_Gamma2_ULapp}), and we assume that SUSY is indeed obeyed when the theory is restricted to one copy (see above). We showed in paper III\cite{tissier11b} that SUSY implies nontrivial WT identities relating cumulants of different orders. More specifically, the following WT identity at the scale $k$ will be needed below:
\begin{equation}
\label{eq_WT_SUSY}
\begin{aligned}
\partial_{1\mu}\Gamma_{k2;x_1,x_2}^{(11)}[\phi,\phi] &- \frac{\Delta_B}{2} (x_1^\mu-x_2^\mu)\Gamma_{k1;x_1x_2}^{(2)}[\phi]  = \\&- \int_{x_3}\phi(x_3)
\partial_{3\mu}\Gamma_{k2;x_1x_3,x_2}^{(21)}[\phi,\phi].
\end{aligned}
\end{equation}

To prove the dimensional reduction property, it is then useful to single out a $2$-dimensional subspace of the $d$-dimensional Euclidean space. We define $x=(y,z)$ with $y\in R^{d-2}$ and $z\in R^2$ and we consider superfields that are uniform in the 2-dimensional Grassmannian and the 2-dimensional Euclidean subspaces, \textit{i.e.} $\Phi(\underline{x})=\phi(y)$. For such fields, the WT identity in Eq.~(\ref{eq_WT_SUSY}) implies that
\begin{equation}
\label{eq_WT_SUSY_vertex}
\Gamma_{k2;y_1,y_2}^{(11)}[p^2;\phi,\phi] = \Delta_B\, \partial_{p^2}\Gamma_{k1;y_1y_2}^{(2)}[p^2;\phi],
\end{equation}
where $p$ is the $2$-dimensional momentum obtained from a spatial Fourier transform over $z$ [by using translational invariance we have as usual defined $\Gamma_{k1}^{(2)}(p_1,p_2) = (2\pi)^2 \delta^{(2)}(p_1+p_2)\Gamma_{k1}^{(2)}(p^2)$ and similarly for $\Gamma_{k2}^{(11)}(p_1,p_2)$]. The same type of relation holds between $- \widetilde{R}_{k;y_1,y_2}(p^2)$ and $\widehat{R}_{k;y_1,y_2}(p^2)$.

With the help of the above relations, the ERGE for the first cumulant $\Gamma_{k1}[\phi]$, Eq.~(\ref{eq_flow_Gamma1_ULapp}), can be reexpressed as
\begin{equation}
\label{eq_gamma1_DR}
\begin{aligned}
\partial_t \Gamma_{k1}[\phi]&=- \big(\frac{\Delta_B}{2} \big) \tilde \partial_t  \int d^{d-2}y_1 \int d^{d-2}y_2 \int \frac{d^{2}p}{(2\pi)^2}\\&\times \widehat{P}_{k;y_1y_2}[p^2, \phi]\; \partial_{p^2}\left( \Gamma_{k1;y_1y_2}^{(2)}[p^2,\phi] +  \widehat{R}_k(p^2)\right) \\&= - \big(\frac{\Delta_B}{2} \big) \tilde \partial_t  \int d^{d-2}y_1 \int \frac{d(p^2)}{4\pi}\\&\times \partial_{p^2}\left[ \log \left( \Gamma_{k1}^{(2)}[p^2,\phi] +  \widehat{R}_k(p^2)\right)\right] _{y_1y_1},
\end{aligned}
\end{equation}
where, in the last expression, $\Gamma_{k1}^{(2)}$ and $\widehat{R}_k$ are functions of $p^2$ but operators in the $(d-2)$-dimensional Euclidean space spanned by the coordinate $y$ (since $\phi(y)$ is \textit{not} uniform). The integral over $p^2$ is easily performed. Choosing an infrared cutoff function that becomes independent of $k$ when its argument $p^2$ is at the UV scale, it only remains:
\begin{equation}
\label{eq_dim_red}
\begin{aligned}
\partial_t \Gamma_{k1}[\phi]=&\big(\frac{\Delta_B}{4\pi} \big) \frac 12 \tilde{\partial}_t \int d^{d-2}y \\&\times \big[ \log\left( \Gamma_{k1}^{(2)}[p^2=0,\phi] +  \widehat{R}_k(p^2=0)\right)\big ]_{yy}.
\end{aligned}
\end{equation}
Up to the trivial factor $\Delta_B/(4\pi)$ and with the identifications
\begin{equation}
\Gamma_{k1}[\phi]\equiv \Gamma_{k}[\phi], \Gamma_{k1}^{(2)}[p^2=0,\phi]\equiv \Gamma_{k}^{(2)}[\phi],  \widehat{R}_k(p^2=0)\equiv R_k,
\end{equation}
Eq.~(\ref{eq_dim_red}) coincides with the ERGE for the effective average action of a standard scalar $\phi^4$ theory\cite{berges02}  in dimension $d-2$. This provides another nonperturbative demonstration\cite{cardy83,klein83,klein84} of the property of dimensional reduction for supersymmetric scalar field theories, first put forward by Parisi and Sourlas.\cite{parisi79}

\subsection{Spontaneous SUSY breaking}

We have shown within a nonperturbative implementation of the RG that the supersymmetry of the theory, and more specifically the superrotational invariance for one copy (SUSY), imply the property of dimensional reduction. As one knows that dimensional reduction does not hold in low enough dimension, what then goes wrong in the formalism?

The answer is that some (super)symmetries or identities must be spontaneously broken along the RG flow. We make the assumption, which will be supported by actual calculations, that only the superrotational invariance (for the theory restricted to a single copy) may be broken along the RG flow, all other (super)symmetries and properties, most significantly the ``Grassmannian ultralocality''  encoding ground-state dominance (when $\beta \rightarrow \infty$), remaining unaltered. From the above proof, it follows that failure of dimensional reduction implies that the WT identity between $\Gamma_{k2}^{(11)}$ and $\Gamma_{k1}^{(2)}$ breaks down: a singularity must occur along the flow, which invalidates the WT identity; the equality in Eq.~(\ref{eq_flow_ward}) and some step in its derivation must go wrong at some scale $k$. From our previous work,\cite{tarjus04,tissier06,tarjus08,tissier08} we anticipate that this arises due to the appearance of a strong enough nonanalytic dependence of the second cumulant of the renormalized random field $\Gamma_{k2}^{(11)}[\phi_1,\phi_2]$ in its field arguments when the latter two, $\phi_1$ and $\phi_2$, become equal. (One should keep in mind that the cumulants are invariant under permutations of their arguments; consequently, $\Gamma_{k2}^{(11)}$ is an even function(al) in $\phi_1 - \phi_2$.) As a result, some higher-order derivative (\textit{i.e.}, a 1PI vertex deriving from $\Gamma_{k2}^{(11)}$) blows up and the whole hierarchy of coupled ERGE's and WT identities for one copy ceases to be valid.

In order to envisage the possible scenario for spontaneous SUSY breaking and breakdown of dimensional reduction, it is instructive to study the structure of the exact hierarchy of coupled RG flow equations, whose lowest-order examples are given in Eqs.~(\ref{eq_flow_Gamma1_ULapp},\ref{eq_flow_Gamma2_ULapp}). The following exposition is not meant to be rigorous, but only to provide some heuristic arguments.

Consider first the hierarchy of ERGE's for the proper vertices evaluated for uniform fields (in Euclidean space) that is obtained by repeated functional differentiation of the ERGE for the effective average action. For this part of the reasoning, the case of a simple scalar field theory is sufficiently illustrative. The ERGE for the effective average action then reads
\begin{equation}
\partial_t \Gamma_k[\phi]=\frac{1}{2}\int_q \partial_t R_k(q^2) \left(\Gamma_k^{(2)}[\phi]+R_k \right)_{q,-q}^{-1}.
\end{equation}
After differentiating this equation $s$ times, it is easily realized that for $s\geq 2$ the ERGE for any generic 1PI vertex $\Gamma_k^{(s)}$  depends on all lower-order proper vertices (of order $\geq 2$) as well as on some $\Gamma_k^{(s+1)}$'s and $\Gamma_k^{(s+2)}$'s. Two cases then arise: when $s\geq 5$, the flow equation is \textit{linear} in $\Gamma_k^{(s)}$ itself, whereas in the other case when $4\geq s\geq2$, the flow equation may be \textit{nonlinear} in $\Gamma_k^{(s)}$ itself. The same structure applies to the present theory. A consequence is that, if no other 1PI vertices diverge first, $\Gamma_k^{(s)}$ can only diverge at the end of the flow ($k\rightarrow 0$) when $s\geq 5$ (due to the linearity of the corresponding ERGE) whereas a divergence may occur at a finite scale when $s\leq 4$.

To get more insight, we have to look more thoroughly into the structure of the RG equations and consider the 1PI vertices associated with the cumulants separately [see  Eqs.~(\ref{eq_flow_Gamma1_ULapp},\ref{eq_flow_Gamma2_ULapp}) for illustration]. We further make the following plausible assumptions (in order to have a renormalizable theory):

(i) The $1$-copy 1PI vertices $\Gamma_{k1}^{(s)}$ always stay finite, as do the cumulants of the renormalized random field $\Gamma_{kp}^{(11...1)}$ and their first-order derivatives. The latter may however be discontinuous; the condition on the first derivatives amounts to assuming that no ``supercusp''\cite{footnote2} stronger than linear appears in the cumulants when two arguments become equal.

(ii) Since we expect a nonanalytic behavior occuring as the copy fields become equal in the reduction to an effective $1$-copy system, we also anticipate that derivatives with respect to symmetric combinations of the field arguments are always bounded (\textit{e.g.}, $\vert(\partial \phi_1 +\partial \phi_2)\Gamma_{kp}^{(s_1s_2...s_p)}[\phi_1,\phi_2,...\phi_p] \vert_{\phi} \vert < +\infty$, where the derivative is evaluated for all replica fields equal to $\phi$).

A ``cusp''\cite{footnote2} in $\Gamma_{kp}^{(11...1)}$ with $p\geq 3$ implies that one of its  second derivatives, which means a proper vertex of order at least $s=5$, blows up. Similarly, for having a ``subcusp''\cite{footnote2} in $\Gamma_{k2}^{(11)}$, at least one of its third derivatives, hence a proper vertex of order at least $s=5$, should blow up. As we have seen above that the associated flow equations are linear, we conclude that the divergence of such proper vertices (with $s\geq 5$) cannot be the triggering event for a singularity at a finite scale. On the other hand, due to the nonlinearity of the ERGE's for the second derivatives of $\Gamma_{k2}^{(11)}$, a ``cusp'' may appear in the latter at a finite scale.

To summarize this discussion: one expects that breakdown of dimensional reduction requires the presence of a  \textit{cusp} in the field dependence of $\Gamma_{k2}^{(11)}$, cusp that should appear \textit{at a finite scale}, which by analogy with what occurs for a manifold in a random environment\cite{larkin70,BLchauve00,CUSPledoussal} we call the ``Larkin'' scale, during the RG flow. By contrast, weaker nonanalyticities in $\Gamma_{k2}^{(11)}$ as well as nonanalyticities in higher-order cumulants (if no cusp has appeared in $\Gamma_{k2}^{(11)}$) can only appear at the fixed point, in the limit $k\rightarrow0$, thereby preserving the dimensional-reduction property. Indeed, the derivation leading to Eq.~(\ref{eq_dim_red}) then remains valid.

\subsection{Continuing the NP-FRG flow with broken SUSY}

Another nontrivial question is then raised: if superrotational invariance (SUSY) is spontaneously broken, \textit{how can one continue the RG flow for the effective average action?} The answer, which again should be verified in calculations, is as follows:

(i) Keeping in mind that superrotational invariance is explicitly broken in the presence of multiple copies (see section~\ref{sec:model}) and that it is only recovered in the process of reducing the whole problem to a single-copy system, spontaneous breaking may then be described by rejecting any implicit assumption of analyticity of the field dependences in the latter process. One should therefore restrict the hierarchy of ERGE's to those equations for cumulants that are considered for generic (nonequal) field arguments, so that a putative nonanalytic dependence in these arguments can freely emerge along the RG flow.\cite{footnote3}

(ii) One assumes that except for the superrotational invariance, all of the properties and symmetries of the effective average action $\Gamma_{k}$ remain valid. In particular, this applies to the property of ``Grassmannian ultralocality" of the random generating functional and to its consequences for the cumulants of the renormalized disorder. As discussed in the preceding paper, this property is obeyed when taking the limit $\beta^{-1}=0$ in the ERGE's for the cumulants, which leads to Eqs.~(\ref{eq_flow_Gamma1_ULapp},\ref{eq_flow_Gamma2_ULapp}) and their extensions to higher orders. It keeps track of the fact that only the ground state is considered for each copy. The validity of the property of ``Grassmannian ultralocality" is therefore distinct from that of superrotational invariance, a distinction that cannot be made in the original Parisi-Sourlas formalism.

(iii) The cutoff functions satisfy the relation in Eq.~(\ref{eq_cutoffs_relation}) which contains as a special case the explicit superrotational invariance when $\Delta_k=\Delta_B Z_k$. In practice, the typical strength of the renormalized random field $\Delta_k$ is defined from $\Gamma_{k2}^{(11)}(q^2=0)$ evaluated for some specific uniform field configuration and the field renormalization constant $Z_k$ is obtained from $\Gamma_{k1}^{(2)}(q^2=0)$ also evaluated for some uniform field configuration.
Finally, the cusp in $\Gamma_{k2;x_1,x_2}^{(11)}(\phi_1,\phi_2)$, which is associated with spontaneous SUSY breaking, must be stable upon further evolution with the infrared scale $k$ and must not generate ``supercusps''. From the structure of the ERGE for $\Gamma_{k2;x_1,x_2}^{(11)}(\phi_1,\phi_2)$, it is expected that a ``linear cusp''\cite{footnote2} satisfies these requirements and provides a mechanism for dimensional reduction \textit{if} it persists at the fixed point (see Appendix \ref{appendixA}).

\section{Approximation scheme}
\label{sec:approximation scheme}

\subsection{SUSY-compatible approximation scheme}

Up to this point, many of the previous considerations remain plausible conjectures. We now provide an implementation of the NP-FRG that allows us to check their validity. The hierachy of ERGE's of course cannot be solved exactly and the next step is to provide a SUSY-compatible nonperturbative approximation scheme. From our previous work,\cite{tarjus04,tissier06,tarjus08,tissier08} we know that an efficient scheme relies on a joint truncation in the \textit{derivative expansion}, which approximates the long-distance behavior of the 1PI vertices,\cite{berges02} and in the \textit{expansion in cumulants of the renormalized disorder}. The truncations however must be combined in a way that does not explicitly break the supersymmetry, more precisely the superrotational invariance (SUSY). To implement this requirement, we use the WT identities.

As is clear from the discussion in the preceding section, a minimal truncation must at least include the second renormalized cumulant. When SUSY is not broken, the WT identity [see Eq.~(\ref{eq_WT_SUSY_vertex})] imposes that for a uniform field $\Gamma_{k2}^{(11)}(q^2;\phi,\phi)$ is given by the derivative of  $\Gamma_{k1}^{(2)}(q^2;\phi)$ with respect to $q^2$. (A similar relation is satisfied by the infrared cutoff functions.) Consider now the derivative expansion. In the lowest order (LPA for ``local potential approximation''), the derivative of $\Gamma_{k1}^{(2)}$ is a field-independent constant, which would force $\Gamma_{k2}^{(11)}$ to be also field independent: this is clearly too crude to describe the physics of the RFIM. The first order of the derivative expansion leads to $\Gamma_{k1}^{(2)}(q^2;\phi)$ of the form  $U_k''(\phi) + Z_k(\phi)q^2$, with $U_k(\phi)$ the ``effective average potential'' (``Gibbs free energy'' for a magnetic system) and $Z_k(\phi)$ a field renormalization function. When the WT identity is satisfied, this in turn requires that
\begin{equation}
\label{eq_WT_approx}
\Gamma_{k2}^{(11)}(q^2;\phi,\phi)=\Delta_B Z_k(\phi),
\end{equation}
which corresponds to a local approximation for the second cumulant. It is easy to generalize this reasoning by taking into account the whole set of WT identities for the proper vertices obtained from the cumulants: a SUSY-compatible approximation at order $n$ consists of taking $\Gamma_{k1}$ at the order $n$ of the derivative expansion, $\Gamma_{k2}$ at the order $n-1$ of the derivative expansion, ..., $\Gamma_{k(n+1)}$ in the local approximation, and all higher-order cumulants equal to zero.\cite{footnote5} This provides a scheme of successive approximations that in principle can be used to check the robustness of the results and if necessary improve them.

The minimal truncation that can already describe the long-distance physics of the RFIM and does not explicitly break SUSY is then the following:
\begin{equation}
\label{eq_ansatz1}
\begin{aligned}
\Gamma_{k1}[\phi]=\int_{x}\left[ U_k(\phi(x))+\frac{1}{2}Z_k(\phi(x))(\partial_{\mu}\phi(x))^2 \right],
\end{aligned}
\end{equation}
\begin{equation}
\label{eq_ansatz2}
\Gamma_{k2}[\phi_1,\phi_2]=\int_{x}V_k(\phi_1(x),\phi_2(x)),
\end{equation}
with the higher-order cumulants set to zero.\cite{footnote51} 

Inserted in the ERGE's for the cumulants, Eqs.~(\ref{eq_flow_Gamma1_ULapp}) and (\ref{eq_flow_Gamma2_ULapp}), the above ansatz provides $3$ coupled flow equations for the $1$-copy effective average potential $U_k(\phi)$ (or its derivative) that describes the thermodynamics of the system, the field renormalization function $Z_k(\phi)$, and the $2$-copy effective average potential $V_k(\phi_1,\phi_2)$ from which one obtains the second cumulant of the renormalized random field at zero momentum, $\Gamma_{k2}^{(11)}(q^2=0;\phi_1,\phi_2) \equiv \Delta_k(\phi_1,\phi_2) =V_k^{(11)}(\phi_1,\phi_2)$. SUSY is obeyed when 
$\Delta_k(\phi,\phi)=Z_k(\phi)$, which is easily satisfied at the UV scale $k=\Lambda$.

\subsection{NP-FRG equations in a scaled form}

In order to search for the fixed point that controls the critical behavior, the flow equations must be recast in a scaled form. This can be done by introducing appropriate scaling dimensions (see refs.~[\onlinecite{tarjus04,tarjus08}] and paper III\cite{tissier11b}). Near a zero-temperature fixed point,\cite{villain84,fisher86} one has the following scaling dimensions: 
\begin{equation}
Z_{k} \sim k^{-\eta}, \;\Delta_k \sim k^{-(2 \eta - \bar \eta)},\;   \phi_a \sim k^{\frac{1}{2}(d-4+\bar \eta)},
\end{equation}
and a renormalized temperature is introduced as $T_k\sim \frac{k^2 Z_k}{\beta \Delta_k} \sim k^\theta/\beta$, with $\theta$ and $\bar \eta$ related through $\theta=2+\eta-\bar \eta$. More precisely, we define running anomalous dimensions $\eta_k$ and $\bar \eta_k$ as
\begin{equation}
  \label{eq_eta_Z}
\partial_t \log Z_{k}= -\eta_k ,
\end{equation}
\begin{equation}
  \label{eq_etabar_Delta}
\partial_t \log \Delta_k =-(2 \eta_k - \bar \eta_k) ,
\end{equation}
where $Z_{k}$ and $\Delta_k$ have been introduced in Sec.~IV-D. One also has
\begin{equation}
U_k\sim \frac{k^d}{T_k}\sim k^{d-\theta}, \;V_k\sim \frac{k^d}{T_k^2}\sim k^{d-2\theta}.
\end{equation}
The dimensionless counterparts of $U_k, V_k,\Delta _k, \phi$ will be denoted by lower-case letters, $u_k, v_k,\delta _k, \varphi$. 

The resulting equations are
\begin{equation}
\begin{split}
  \label{eq_u'_ising}
&\partial_t u_k'
  (\varphi)=-\frac{1}{2}(d-2\eta_k+\bar\eta_k)u_k'(\varphi) +
  \frac{1}{2}(d-4+\bar\eta_k) \varphi\; u_k''(\varphi)\\& +
  2v_d\Big\{-\frac{d-2}{2} l_1^{(d-2)}(\varphi)\,u'''_k(\varphi) +
  \big[ l_2^{(d+2)}(\varphi) z_{k}'(\varphi) + l_2^{(d)}(\varphi)
  \\& \times u'''_k (\varphi) \big] \big[ z_{k}(\varphi)
  -\delta_{k,0}(\varphi)\big]  + l_1^{(d)}(\varphi)\big[-\frac{d}{2}
  z_k'(\varphi) + \delta_{k,0}'(\varphi)\big]\\&+\frac{1}{2}(\eta_k-\bar\eta_k)\big[z_k'(\varphi)j_2^{(1,d+2)}(\varphi)+u'''_k(\varphi)j_2^{(1,d)}(\varphi)\big] \Big\},
\end{split}
\end{equation}

\begin{equation}
 \label{eq_z_ising}
\begin{split}
&\partial_t z_k(\varphi)= \eta_k z_{k}(\varphi) +  \frac{1}{2}(d-4+\bar\eta_k) \varphi\; z_{k}'(\varphi) -2 v_d \Big\{\frac{d-2}{2} \times \\&  l_{1}^{(d-2)}(\varphi) z_{k}''(\varphi) - (d-2) l_{2}^{(d-2)}(\varphi) z_{k}'(\varphi)u'''_k(\varphi) + 2\frac{2d+1}{d}\times \\& l_{3}^{(d+2)}(\varphi) z_{k}'(\varphi)^2 [z_{k}(\varphi) - \delta_{k,0}(\varphi)] + 4 l_{3}^{(d)}(\varphi) z_{k}'(\varphi) u'''_k(\varphi)[z_{k}(\varphi) \\&- \delta_{k,0}(\varphi)] - l_{2}^{(d)}(\varphi)
\big[\frac{2d+1}{2} z_{k}'(\varphi)^2 + z_{k}''(\varphi)(z_{k}(\varphi) - \delta_{k,0}(\varphi)) \\& - 2 z_{k}'(\varphi) \delta_{k,0}'(\varphi)\big] + m_{4}^{(d-2)}(\varphi)u'''_k(\varphi)^2 + \frac{2}{d}m_{4}^{(d)}(\varphi) u'''_k(\varphi)\times \\&[(d+2)z_{k}'(\varphi)- 2 \delta_{k,0}'(\varphi)] + \frac{1}{d}m_{4}^{(d+2)}(\varphi)z'_k(\varphi)[(d+4)z_{k}'(\varphi)- \\& 4 \delta_{k,0}'(\varphi)] - \frac{8}{d}[m_{5}^{(d+4)}(\varphi) z'_k(\varphi)^2 + 2 m_{5}^{(d+2)}(\varphi) z'_k(\varphi)u'''_k(\varphi) +\\& m_{5}^{(d)}(\varphi) u'''_k(\varphi)^2][z_{k}(\varphi) - \delta_{k,0}(\varphi)]
-\frac{1}{2}(\eta_k-\bar\eta_k)\big[z_k''(\varphi)\times\\&j_2^{(1,d)}(\varphi)-(4+\frac
2d)z_k'(\varphi)^2j_3^{(1,d+2)}(\varphi)-4z_k'(\varphi)u'''_k(\varphi)\times\\&
j_3^{(1,d)}(\varphi)+\frac 2dz_k'(\varphi)^2h_4^{(d+2)}(\varphi)+\frac 4dz_k'(\varphi)u'''_k(\varphi)h_4^{(d)}(\varphi)+\\&
\frac 2d u'''_k(\varphi)^2h_4^{(d-2)}(\varphi)\big] \Big\},
\end{split}
\end{equation}
\begin{equation}
 \label{eq_v_ising}
\begin{split}
&\partial_t v_k(\varphi_1,\varphi_2)=-(d-4+2\bar\eta_k-2\eta_k)  v_k(\varphi_1,\varphi_2) + \\& \frac{1}{2} (d-4+\bar\eta_k) (\varphi_1\partial_{\varphi_1}+\varphi_2\partial_{\varphi_2}) v_k(\varphi_1,\varphi_2) - 2v_d\Big\{\frac{1}{2}\times \\&
l_{1,1}^{(d)}(\varphi_1,\varphi_2)[\delta_k(\varphi_1,\varphi_2)-z_k(\varphi_1)]^2 +
l_{2}^{(d)}(\varphi_1)v_k^{(20)}(\varphi_1,\varphi_2)\\& \times [\delta_{k,0}(\varphi_1)-z_k(\varphi_1)] +  \frac{d-2}{2} l_{1}^{(d-2)}(\varphi_1)v_k^{(20)}(\varphi_1,\varphi_2) + \\& \frac{1}{2} m_{1,1}^{(d-2)}(\varphi_1,\varphi_2) - n_{1,1}^{(d-2)}(\varphi_1,\varphi_2)[\delta_k(\varphi_1,\varphi_2)-z_k(\varphi_1)] \\&
-\frac{1}{2}(\eta_k-\bar\eta_k) \big[\delta_k(\varphi_1,\varphi_2)j_{1,1}^{(1,d)}(\varphi_1,\varphi_2)+j_{1,1}^{(2,d)}(\varphi_1,\varphi_2)\\&+v_k^{(20)}(\varphi_1,\varphi_2)j_{2}^{(1,d)}(\varphi_1)\big] +perm(12) \Big\},
\end{split}
\end{equation}
where  $v_d^{-1}=2^{d+1}\pi^{d/2} \Gamma(d/2)$, a prime denotes a derivative with respect to the field (when only one argument is present), $\delta_k(\varphi_1,\varphi_2)=v_k^{(11)}\left(\varphi_1,  \varphi_2 \right)$, $\delta_{k,0}(\varphi_1) = \delta_k(\varphi_1,\varphi_1)$, and $perm(12)$ denotes the terms obtained by permuting $\varphi_1$ and $\varphi_2$.

The functions $l_{q}^{(d)}(\varphi)$,
$l_{q_1,q_2}^{(d)}(\varphi_1,\varphi_2)$, $m_{q}^{(d)}(\varphi)$,
$m_{q_1,q_2}^{(d)}(\varphi_1,\varphi_2)$,
$n_{q_1,q_2}^{(d)}(\varphi_1,\varphi_2)$, $h_{q}^{(d)}(\varphi)$,
 $j_{q_1}^{(a,d)}(\varphi)$ and
$j_{q_1,q_2}^{(a,d)}(\varphi_1,\varphi_2)$ appearing in the above flow equations are ``dimensionless threshold functions'' which are defined from the infrared cutoff functions that satisfy\cite{footnote4}
\begin{equation}
\label{eq_scaled_hatR}
\widehat{R}_k(q^2)=Z_k k^2 s(q^2/k^2)
\end{equation}
and
\begin{equation}
\label{eq_scaled_wideR}
\widetilde{R}_k(q^2)=-\Delta_k\, s'(q^2/k^2),
\end{equation}
with the prime again denoting a derivative. The dependence of  the dimensionless threshold functions on the dimensionless fields  comes through $u''_k(\varphi)$ and $z_k(\varphi)$. Their definitions are given in Appendix \ref{appendixB}. The properties of these threshold functions have been extensively discussed.\cite{berges02,tetradis94,delamotte03} They decay rapidly when their arguments $u''_k(\varphi_a)$ become large, which, since $u_k''(\varphi_a)=U_k''(\phi_a)/(Z_k k^2)$ is the square of a renormalized mass, ensures that only modes with a mass smaller than $k$ contribute to the flow in Eqs.~(\ref{eq_u'_ising}-\ref{eq_v_ising}). These threshold functions essentially encode the nonperturbative effects beyond the standard one-loop approximation.\cite{berges02,tetradis94,delamotte03}

Note that $u_k(\varphi)$ and $z_k(\varphi)$ are even functions of $\varphi$ and that, due to the $Z_2$ symmetry and the permutational symmetry of the cumulants in their field arguments, $v_k(\varphi_1,\varphi_2)$ satisfies $v_k(\varphi_1,\varphi_2)=v_k(-\varphi_1,-\varphi_2)=v_k(\varphi_2,\varphi_1)=v_k(-\varphi_2,-\varphi_1)$, and similarly for $\delta_k(\varphi_1,\varphi_2)$. Finally, the initial conditions for Eqs.~(\ref{eq_eta_Z},\ref{eq_etabar_Delta},\ref{eq_u'_ising},\ref{eq_z_ising},\ref{eq_v_ising}) at the UV scale $\Lambda$ ($t=0$) are obtained from the bare action in Eq.~(\ref{eq_superaction_multicopy}) for $\beta=0$ and $\Phi_a=\varphi_a$, \textit{i.e.},
\begin{align}
& z_{\Lambda}(\varphi)=1\\&
u_{\Lambda}(\varphi)=\frac{\tau}{2}\varphi^2+\frac{u}{4!}\varphi^4=\frac{u}{4!}(\varphi^2-\kappa_{\Lambda})^2,
\end{align}
the last equality being valid in the (relevant) region where the bare potential has a nontrivial minimum;\cite{berges02}  after a trivial rescaling of the fields so that the explicit dependence on $\Delta_B$ disappears, one also has
\begin{equation}
v_{\Lambda}(\varphi_1,\varphi_2) = \varphi_1\varphi_2,
\end{equation}
which implies $\delta_{\Lambda}(\varphi_1,\varphi_2) =1$. This initial
condition trivially satisfies $\delta_{\Lambda}(\varphi,\varphi) =
z_{\Lambda}(\varphi)$ which ensures that  invariance under superrotations is obeyed at the beginning of the
flow.

\subsection{Properties of the flow equations}

An important property of the present theory occurs in the limit $\varphi_2 \rightarrow \varphi_1$ when $\delta_k(\varphi_1,\varphi_2)$ is regular enough. For convenience, we introduce the variables $\varphi=(\varphi_1+\varphi_2)/2$ and $y=(\varphi_1-\varphi_2)/2$. The property can be stated as follows: If the behavior of $\delta_k$ is regular enough when $y\rightarrow 0$, \textit{i.e.}, is such that
\begin{equation}
 \label{eq_deltak_small_y}
\delta_k(\varphi,y)=\delta_{k,0}(\varphi) +\frac{1}{2}\delta_{k,2}(\varphi)y^2 +y^2\,o(y)
\end{equation}
with $o(y)\rightarrow 0$ when $y\rightarrow 0$, then the flow of $\delta_{k,0}(\varphi)$ coincides with that of $z_k(\varphi)$.  This is precisely the WT relation in Eq.~(\ref{eq_WT_SUSY_vertex}) when evaluated at zero momentum; indeed, in dimensionless form, this equation is just $\delta_{k,0}(\varphi)=z_k(\varphi)$. Within the present ansatz for the effective average action, dimensional reduction then follows.

To prove the equality of the two flow equations, we first derive Eq.~(\ref{eq_v_ising}) with respect to $\varphi_1$ and $\varphi_2$ to obtain the RG equation for $\delta_k$, change variables to $\varphi,y$ and take the limit $y=0$ by assuming the above small $y$ behavior [Eq.~(\ref{eq_deltak_small_y})]. The derivation makes use of the properties of the dimensionless threhold functions that are summarized in Appendix \ref{appendixB}. This leads to
\begin{equation}
 \label{eq_delta0_ising}
\begin{split}
&\partial_t \delta_{k,0}(\varphi)=-(\bar\eta_k-2\eta_k) \delta_{k,0}(\varphi)  + \frac{1}{2} (d-4+\bar\eta_k) \varphi \delta_{k,0}'(\varphi) \\& -2 v_d \Big\{\frac{d-2}{2} \left[  l_{1}^{(d-2)}(\varphi) \delta_{k,0}''(\varphi) - 2 l_{2}^{(d-2)}(\varphi)u'''_k(\varphi) \delta_{k,0}'(\varphi)\right] \\& - \frac{2d-3}{2} l_{2}^{(d)}(\varphi) \delta_{k,0}'(\varphi)^2  + m_{4}^{(d-2)}(\varphi)u'''_k(\varphi)^2 + 2 m_{4}^{(d)}(\varphi) \times \\&u'''_k(\varphi)\delta_{k,0}'(\varphi) + m_{4}^{(d+2)}(\varphi)\delta_{k,0}'(\varphi)^2+ 
L_k[\varphi;\delta_{k,0}-z_k]\Big\},
\end{split}
\end{equation}
where $L_k[\varphi;\delta_{k,0}-z_k]$ is a polynomial of degree two in $[\delta_{k,0}(\varphi)-z_k(\varphi)]$ and its derivatives, with coefficients that are function of $\varphi$ and such that $L_k[\varphi;0]\equiv 0$. Subtracting from the above flow equation the one for $z_k$ in Eq.~(\ref{eq_z_ising}) leads to an equation for the difference $\delta_{k,0}-z_k$ in the form
\begin{equation}
 \label{eq_diff_ising}
\partial_t \left[ \delta_{k,0}(\varphi)-z_k(\varphi) \right] = M_k[\varphi; \delta_{k,0}-z_k] - (\bar\eta_k-\eta_k) z_k(\varphi),
\end{equation}
where $M_k[\varphi; \delta_{k,0}-z_k]$ is another  polynomial of degree two in $[\delta_{k,0}(\varphi)-z_k(\varphi)]$ and its derivatives such that $M_k[\varphi;0]\equiv 0$. [Note that due to their definitions, $\bar\eta_k$ and $\eta_k$ are equal as soon as $\delta_{k,0}(\varphi)=z_k(\varphi)$.] Since the initial condition satisfies $\delta_{k=\Lambda}(\varphi_1,\varphi_2)\equiv 1$ and $z_{k=\Lambda}(\varphi)\equiv 1$, the only solution to Eq.~(\ref{eq_diff_ising}) is $\delta_{k,0}(\varphi)-z_k(\varphi)=0$ at all scales, which provides the desired demonstration. By inserting this result in Eq.~(\ref{eq_u'_ising}), one obtains 
\begin{equation}
\begin{split}
  \label{eq_u'_ising_DR}
&\partial_t u_k' (\varphi)=-\frac{1}{2}(d-\eta_k)u_k'(\varphi) + \frac{1}{2}(d-4+\eta_k) \varphi\; u_k''(\varphi)\\& - (d-2) v_d\Big\{ l_1^{(d-2)}(\varphi)\,u'''_k(\varphi) + l_1^{(d)}(\varphi)\, z_k'(\varphi) \Big\},
\end{split}
\end{equation}
which coincides with the flow equation for $u'_k(\varphi)$ in the $\varphi^4$ theory without random field in dimension $d-2$ (use the relation $(d-2)v_d=(2\pi)^{-1}v_{d-2}$ and compare with Eq.~(4.33) of Ref.~[\onlinecite{berges02}]). The same exercise can be repeated for the flow of $z_k(\varphi)$, thereby completing the proof of dimensional reduction in this nonperturbative approximation.

On the other hand, a spontaneous breaking of SUSY and of the associated WT identity occurs whenever $\delta_{k,2}(\varphi)$ diverges and $\delta_k$ has a cusp-like singularity in the form
\begin{equation}
\label{eq_cusp}
\delta_k(\varphi_1,\varphi_2)=\delta_{k,0}(\varphi) +\delta_{k,1}(\varphi)|\varphi_1 - \varphi_2| +\cdots
\end{equation}
as $\varphi_2  \rightarrow \varphi_1 \rightarrow \varphi$. We expect that such a cusp appears at a finite ``Larkin'' scale $k_L$  through the formation of a boundary layer as $k\rightarrow k_L$ and $y\rightarrow 0$ in the form
\begin{equation}
 \label{eq_boundary_layer}
\partial^{2}y\;\delta_{k}(\varphi,y) \sim \frac{1}{[k-k_L(\varphi)]^{2\varpi} + y^2},
\end{equation}
with $\varpi$ an (\textit{a priori} unknown) exponent, as can be checked by inserting in the flow equations. The boundary layer provides the appropriate initial conditions to further continue the flow below $k_L$ with a cuspy function.

Finally, it is worth comparing the above NP-FRG equations with those we have previously derived  without the superfield formalism\cite{tarjus08,tissier08} in a minimal approximation based on the same ansatz as in Eqs.~(\ref{eq_ansatz1},\ref{eq_ansatz2}). The two main differences are that:

(i) The present equations are directly considered at zero temperature (and zero auxiliary temperature) and subdominant terms proportional to $T_k$ do not appear in the beta functionals.

(ii) More importantly, since from the previous formalism we had no insight into the requirements for avoiding an \textit{explicit} breaking of the underlying superrotational invariance at the single-copy level, we had for simplicity set the infrared cutoff function $\widetilde R_k$ to zero. As a result, the WT identity associated with superrotational invariance was violated at \textit{all} scales. The dimensional-reduction property could therefore not be exactly recovered when it was expected to be valid.\cite{tarjus08,tissier08}

\subsection{Numerical resolution}

We are interested in the long-distance physics of the RFIM near its critical point and in the associated (once unstable)
fixed point. Such information can be obtained from the solution of the
above coupled FRG flow equations with appropriate boundary
conditions. We have carried out the three following types of resolution.

(1) We have investigated the location of the breakdown of
SUSY associated with the appearance of a cusp in
$\delta_k(\varphi_1,\varphi_2)$. To this end, we have solved the set
of coupled flow equations which is obtained when $\delta_k$ has a
sufficiently regular (no cusp) behavior as $\varphi_1\rightarrow
\varphi_2$, \textit{i.e.} $y\rightarrow 0$: see
Eq.~(\ref{eq_deltak_small_y}). As already shown, the flow of
$\delta_{k,0}(\varphi)$ then coincides with that of $z_k(\varphi)$,
and the NP-FRG equations for $u_k'(\varphi)$ and $z_k(\varphi)$ form a
closed set. In addition, we have solved the equation for the second
derivative of $\delta_{k}$ with respect to $(\varphi_1-\varphi_2)$,
$\delta_{k,2}(\varphi)$, which depends on $u_k'(\varphi)$ and
$z_k(\varphi)$. A divergence in $\delta_{k,2}(\varphi)$ signals the
appearance of a cusp in $\delta_k(\varphi_1,\varphi_2)$. 

The numerical task of solving this system of three coupled partial
differential equation is reasonable. We discretize the problem both
in the time and field directions. We use finite differences to
evaluate the field derivatives that appear in the beta functions and
use an explicit scheme to find the (RG) ``time'' evolution of the functions. 

(2) When a cusp is present, solving in their full glory the set of
coupled partial differential equations for functions of up to two
field arguments is a hard task. We have discretized the field
variables, using a 2-dimensional grid to solve the equation for
$\delta_k(\varphi_1,\varphi_2)$. For convenience, we have switched
variables from $\varphi_1,\varphi_2$ to $X=\varphi_1 \varphi_2/2$ and
$y=(\varphi_1-\varphi_2)/2$ and restricted the grid to a trapezoidal
region with $X\geq 0$ and $y\geq 0$. Observe that the flow of
$\delta_k(\varphi_1,\varphi_2)$ depends on 
$\delta_k(\varphi_1,\varphi_1)$ and $\delta_k(\varphi_2,\varphi_2)$
and their derivatives. In this sense, the flow equation for $\delta_k$
is nonlocal since its evolution depends not only on its property in a
neighborhood of $(\varphi_1,\varphi_2)$ but also on its values at
points faraway in the $(\varphi_1,\varphi_2)$ plane. It is therefore
important to choose the trapezoidal grid such that, for all point in
the grid, the information necessary for computing the evolution of the
function at this point is inside the trapezoid. In the larger
simulations we had 60 points in the $X$ direction and 50 in the $y$
direction.

We have solved the set of coupled equations with two kinds of
\textit{modus operandi}:

(2-i) First, for selected values of the dimension
$d$ we have searched the ``cuspy'' fixed point by dichotomy, starting from
``cuspy'' initial conditions ($\delta_{k=\Lambda,1}(x)\neq 0$). Depending
on the dimension and the regulator parameters, there appears from time
to time numerical instabilities. Such instabilities showed up for
instance in the (physically not very interesting) large-field region,
where the threshold functions are very small. To keep the latter under control, we have therefore used slightly modified flow equations: this is discussed in Appendix~\ref{appendix C}. 

(2-ii) Secondly, once a proper ``cuspy'' fixed point has been identified in some
given $d$, we have followed this fixed point by continuously
decreasing or increasing $d$. This was done by looking directly for the 
roots of the beta functions by means of the Hybrid algorithm, using
as an initial guess the fixed point found at the previous iteration. We have studied the cuspy fixed point down to $d\lesssim3$; it  
is indeed difficult to go down to lower $d$ because the anomalous dimensions
$\eta$ and $\bar \eta$ become large in low dimensions: $\bar \eta \simeq
4-d$ (see also Ref.~[\onlinecite{ballhausen04}]).

(3) When approaching the critical dimension $d_{DR}$ for both SUSY and dimensional-reduction breaking from below, the procedures (2-i) and (2-ii) above become inefficient. We have used instead an expansion of the function $\delta_k(\varphi_1=x+y,\varphi_2=x-y)$ in $\vert y\vert$ with, in practice,
\begin{equation}
\delta_k(x+y,x-y)=\delta_{k,0}(x) + \vert y\vert \delta_{k,1}(x) +\frac{y^2}{2}\delta_{k,2}(x) + O(\vert y\vert^3).
\end{equation}
Note that the expansion cannot be applied in the close vicinity of $d_{DR}$. Indeed, one then expects the approach $d \rightarrow d_{DR}^-$ to proceed nonuniformly in $y$ through a boundary layer in $y^2/(d_{DR}-d)$. It is easy to check that a similar boundary-layer phenomenon is found in the more easily handled 1-loop equation for the RF$O(N)$M near $d=4$,\cite{tissier06b} around the critical value $N_{DR}=18$, \textit{i.e.} when $N \rightarrow 18^-$.

To compute the anomalous dimensions, we have defined the amplitudes $Z_k$ and $\Delta_k$ appearing in Eqs.~(\ref{eq_eta_Z},\ref{eq_etabar_Delta}) by choosing the specific configuration of the fields as equal to zero. This amounts to imposing
\begin{equation}
\partial_t z_k(\varphi=0)=0,
\end{equation}
\begin{equation}
 \partial_t \delta_{k,0}(\varphi)= \partial_t \delta_{k}(\varphi,\varphi)=0,
\end{equation}
which fixes $\eta_k$ and $\bar \eta_k$.
 
In addition to the anomalous dimensions
$\eta$ and $\bar \eta$ (obtained as the fixed-point values $\eta_*$ and $\bar \eta_*$), we have computed the critical exponent $\nu$ that controls the escape of the RG flow away from the fixed point along
the unstable (relevant) direction. This was done by diagonalizing the
flow equations around the fixed point and extracting the negative
eigenvalue which identifies with $-1/\nu$.

Finally, we have chosen a dimensionless cutoff function $s(q^2/k^2)$ 
[see Eqs.~(\ref{eq_scaled_hatR}) and (\ref{eq_scaled_wideR})] of the form
\begin{equation}
\label{eq_chosen_r}
s(y)=\left(a+b y +c y^2 \right) e^{-y},
\end{equation}
where $y=q^2/k^2$ [not to be confused with $(\varphi_2 -\varphi_1)/2$]. This function satisfies the requirements $s(y\rightarrow \infty) \rightarrow 0$ and $s(y\rightarrow 0)= \rm{cst}>0$. It is a modification of the function used in Refs.~[\onlinecite{wetterich93},\onlinecite{berges02}]. The parameters $a$, $b$ and $c$ can be optimized via stability considerations (see Refs.~[\onlinecite{litim00},\onlinecite{canet03},\onlinecite{pawlowski07}]). They are varied to find a region of values for which the computed critical exponents are stable under changes of the parameters. For more details, see Appendix~\ref{appendix C}.

\section{Results and discussion}
\label{sec:results}

\subsection{Results}

Our main result is a numerical confirmation of the scenario put
forward above: the NP-FRG equations allowing us to continuously follow
the critical behavior and the associated fixed point as a function of
dimension $d$, we find that there is a critical dimension $d_{DR}$
separating a region $d> d_{DR}$ where SUSY is valid all along the RG
flow (except possibly right at the fixed point) and dimensional reduction
applies from a region $d<d_{DR}$ where SUSY is broken at a finite IR
scale along the RG flow and a breakdown of dimensional reduction takes
place. By using the procedure (1) detailed in the above subsection, we
have numerically located this critical dimension as $d_{DR}\simeq 5.1
\pm 0.1$ (the precise value has a residual dependence on the chosen
cutoff function). Note that the value $d_{DR}\simeq 5.1$ obtained here
is consistent with the value found in our previous, and somewhat
cruder, NP-FRG approach of the RF$O(N)$M when extrapolating the
transition line $d_{DR}(N)$ down to $N=1$ (see Figure 4 of paper
II\cite{tissier08}).

\begin{figure}[ht]
\includegraphics[width=\linewidth]{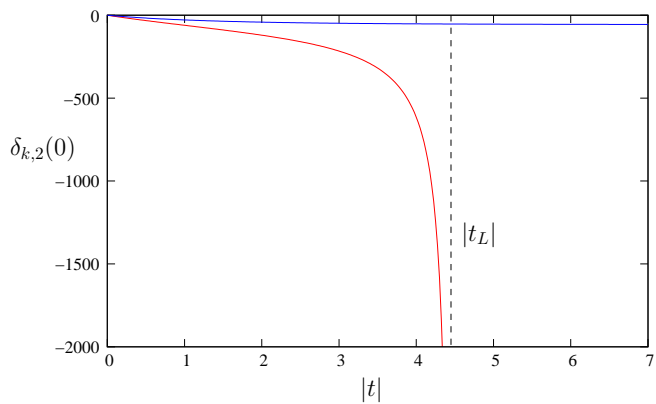}
\caption{\label{fig_larkin} NP-FRG flow of $\delta_{k,2}(0)$ in the regime where SUSY is valid. The initial conditions at $k=\Lambda$ (\textit{i.e.}, $t=0$) for $u'_k(\rho)$ and $z_k(\rho)=\delta_{k,0}(\rho)$, with $\rho=\varphi^2/2$, are taken at the fixed-point solution, $u'_{*}(\rho)$ and $z_{*}(\rho)$ [$\partial_t u_k'(\rho)\vert_{*}=\partial_t z_k(\rho)\vert_{*}=0$], and those for $\delta_{k,0}(\rho)$ and $\delta_{k,2}(\rho)$ are chosen as  $\delta_{k=\Lambda,0}(\rho)=z_{*}(\rho)$ and $\delta_{k=\Lambda,2}(\rho)=0$. The upper (color online blue) curve corresponds to $d=5.2>d_{DR}$ and one observes that  $\delta_{k,2}(0)$ tends to a finite fixed-point value. The lower  (color online red) curve corresponds to $d=5<d_{DR}$ shows a divergence at a finite RG ``time'' $t_L$.}
\end{figure}

\begin{figure}[ht]
\includegraphics[width=\linewidth]{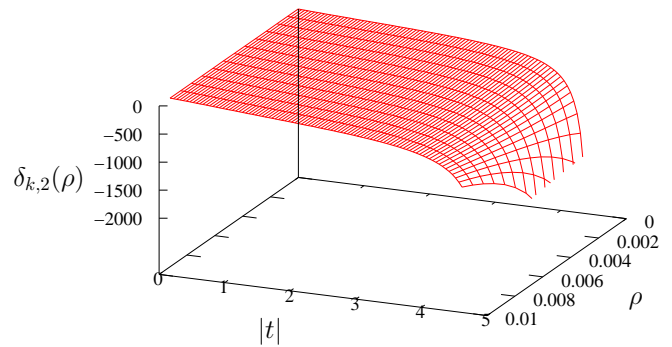}
\caption{\label{fig_larkin_3D} NP-FRG flow of $\delta_{k,2}(\rho)=\partial^2_y\delta(\rho,y)|_{y=0}$ for $d=5<d_{DR}$. The initial conditions at $k=\Lambda$ (\textit{i.e.}, $t=0$) for $u'_k(\rho)$ and $z_k(\rho)$ are taken at the fixed-point solution, $u'_{*}(\rho)$ and $z_{*}(\rho)$, and those for $\delta_{k,0}(\rho)$ and $\delta_{k,2}(\rho)$ are chosen as  $\delta_{k=\Lambda,0}(\rho)=z_{*}(\rho)$ and $\delta_{k=\Lambda,2}(\rho)=0$. One observes that the divergence  first takes place for small values of $\rho$. }
\end{figure}

\begin{figure}[ht]
\includegraphics[width=\linewidth]{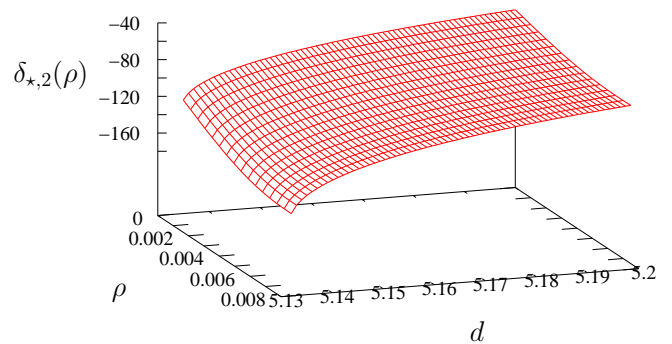}
\caption{\label{fig_d_dr_dpf} Fixed-point solution $\delta_{*,2}(\rho)$ for dimensions $d>d_{DR}$. Note the square-root behavior $\sqrt{d-d_{DR}}$ as a function of dimension.}
\end{figure}

For initial conditions of the RG flow at or near the critical point,
the second derivative $\delta_{k,2}(\varphi)=\partial^2_y
\delta_{k}(\varphi+y,\varphi-y)\vert_{y=0}$ blows up at a finite RG
``time'' $t_L=\log(k_L/\Lambda)$ for $d< d_{DR}$, whereas it stays
finite up to the fixed point for $d> d_{DR}$.\cite{footnote6} We illustrate the
difference of behavior between these two cases in
Fig.~\ref{fig_larkin} for a field configuration $\varphi=0 $.
The divergence of the full function $\delta_{k,2}(\varphi)$ when $d<d_{DR}$ is shown in Fig.~\ref{fig_larkin_3D}.  (Due to the $Z_2$ symmetry, it is more convenient to represent the functions in terms of $\rho=\varphi^2/2$.) 
We find, as can be anticipated from an
analysis of the NP-FRG equations for $u_{k}'(\varphi)$,
$\delta_{k,0}(\varphi)$ and $\delta_{k,2}(\varphi)$, that the latter
stays finite at $d=d_{DR}$ and that its fixed-point value for $d
\rightarrow d_{DR}^+$ behaves as $\delta_{*,2}(\varphi)\vert_{d} -
\delta_{*,2}(\varphi)\vert_{d_{DR}} \propto \sqrt{d-d_{DR}}$, as seen
in Fig.~\ref{fig_d_dr_dpf}.

\begin{figure}[ht]
\includegraphics[width=\linewidth]{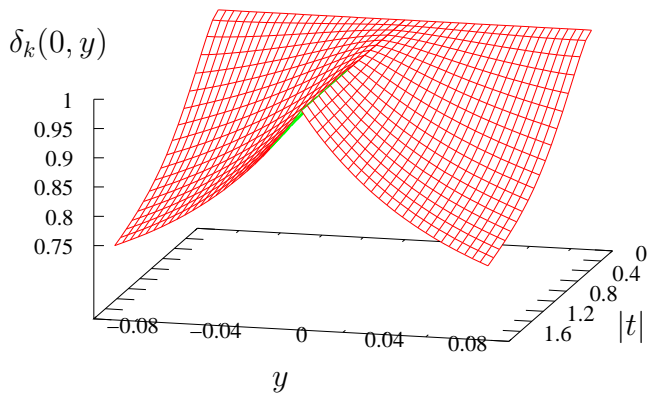}
\caption{\label{fig_larkin_y} NP-FRG flow of $\delta_k(0,y)$ for $d=4<d_{DR}$. The initial conditions at $k=\Lambda$ (\textit{i.e.}, $t=0$) for $u'_k(\rho)$ and $z_k(\rho)$ are taken at the fixed-point solution, $u'_{*}(\rho)$ and $z_{*}(\rho)$,  and that for $\delta_{k}$ is chosen as $\delta_{k=\Lambda}(\rho,y)=1$. One observes that a linear cusp appears at a finite RG time. By construction $\delta_k(0,0)=1$ along the  flow.}
\end{figure}

When considering dimensions smaller than $d_{DR}$, one must study the full
dependence of the function $\delta_k(\rho=\varphi^2/2,y)$. We show in Fig.~\ref{fig_larkin_y} the
evolution of this function for $\rho=0$. Starting from a constant
function, one can clearly see that a linear cusp in $y$ appears in a finite RG time $\vert t_L\vert$, close
to 0.7 for the case shown.

\begin{figure}[ht]
\includegraphics[width=\linewidth]{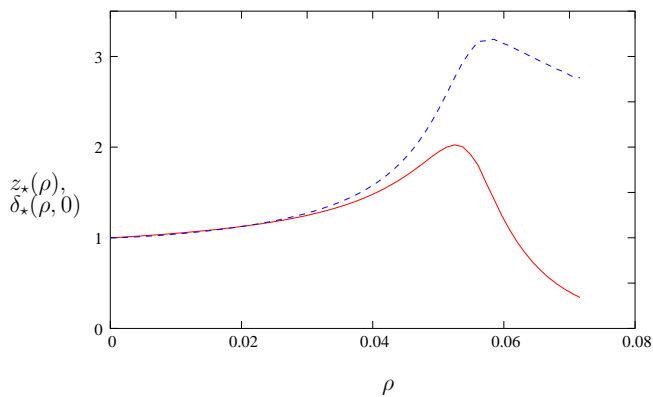}
\caption{\label{fig_d0z3d} Fixed-point solution in $d=3$ for $z_{*}(\rho)$
  (solid line) and $\delta_{*}(\rho,0)$ (dashed line). The two
  functions differ for a large enough field (by construction, they
  coincide at $\rho=0$).}
\end{figure}

The appearance of a cusp along the flow leads to a breakdown of the superrotational invariance and of the associated WT identities.  This is illustrated for the $d=3$ fixed point in Fig~\ref{fig_d0z3d}: there,
$z_{*}(\rho)\neq \delta_{*,0}(\rho) \equiv \delta_{*}(\rho,0)$, which implies a breaking of the WT identity in Eq.~(\ref{eq_WT_approx}).
The different asymptotic behaviors at large $\rho$ are easily deduced
from Eqs.~(\ref{eq_z_ising},\ref{eq_v_ising}), from which we show
that $z_{*}(\rho)\sim \rho^{-\eta/(d-4+\bar\eta)}$ and $\delta_{*}(\rho,0)\sim\rho^{-(2\eta-\bar\eta)/(d-4+\bar\eta)}$. In $d=3$, $2\eta-\bar\eta$ is very small, which implies that the function  $\delta_{*}(\rho,0)$ decreases slowly to $0$.

\begin{figure}[ht]
\includegraphics[width=\linewidth]{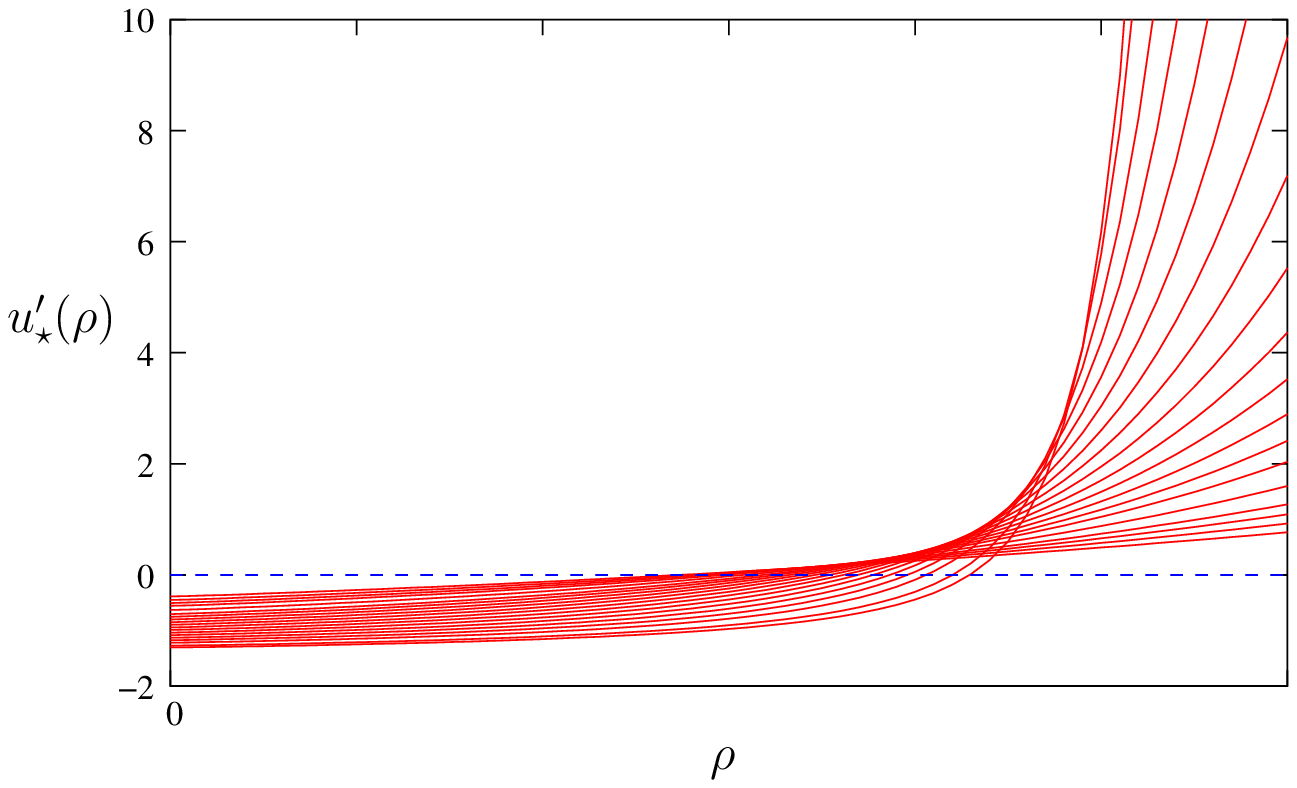}
\caption{\label{fig_uppf} Fixed-point solution $u_{*}'(\rho)$ for dimensions ranging from 3 (steepest) to 4.8 (smoother curve). The  variable $\rho$ has been rescaled by a factor $e^{-d}$ so that all curves can be represented on the same plot. }
\end{figure}
\begin{figure}[ht]
\includegraphics[width=\linewidth]{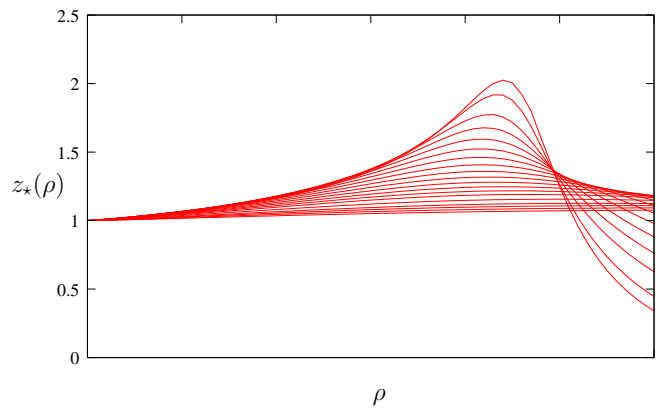}
\caption{\label{fig_zpf} Fixed-point solution $z_{*}(\rho)$ for dimensions
  ranging from 3 (steepest) to 4.8 (smoother curve). The variable
  $\rho$ has been rescaled by a factor $e^{-d}$ so that all curves can
  be represented on the same plot.  }
\end{figure}
\begin{figure}[ht]
\includegraphics[width=\linewidth]{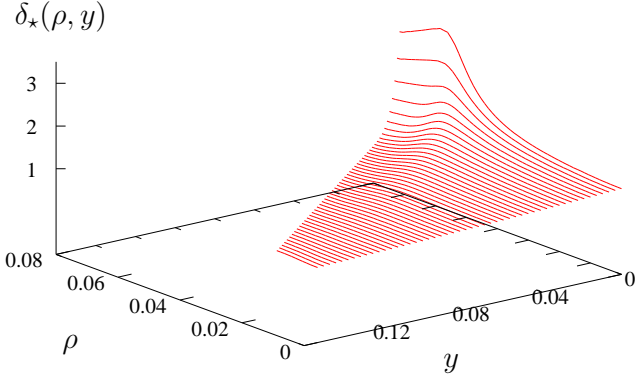}
\caption{\label{fig_deltapf} Fixed-point solution $\delta_{*}(\rho,y)$
  in $d=3$. Note that the point $(\rho=0, y=0)$ corresponds to the
  lower right corner. The function is even in $y$ (with a cusp around
  $y=0$) and the part corresponding to $y<0$ is not shown.}
\end{figure}

We also display in Figs.~\ref{fig_uppf} and~\ref{fig_zpf} the fixed-point
solutions  $u_{*}'(\rho)$ and $z_{*}(\rho)$ for different dimensions. One can see that the lower the dimension the steeper the curves. This means that for a numerical study of the critical properties in low dimensions, we
need to discretize the field dependence in the NP-FRG equations with a small mesh, which entails a large number of 
points. The numerical integration is therefore more difficult and even becomes intractable in practice. Finally, we show in Fig.~\ref{fig_deltapf} the full dependence of the fixed-point solution  $\delta_{*}(\rho,y)$.

\begin{figure}[ht]
\includegraphics[width=\linewidth]{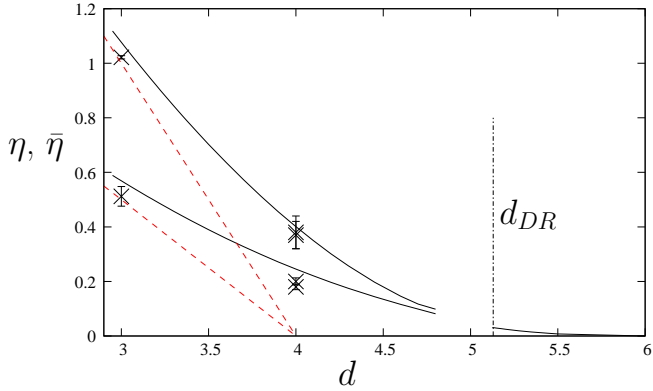}
\caption{\label{fig_eta} Anomalous dimensions $\eta$ and  $\bar\eta$ as functions of the spatial dimension $d$. The dashed lines are lower bounds for the anomalous dimensions [$(4-d)/2$ and $4-d$ for $\eta$ and $\bar \eta$ respectively] and the crosses correspond to predictions from computer simulations and ground-state determinations. }
\end{figure}

We now turn to the results concerning the critical exponents. We begin with the anomalous dimensions $\eta$ and $\bar
\eta$ which are determined at the (critical) fixed point. (We recall that $\eta$ characterizes the spatial decay of the ``connected'' pair correlation function and $\bar \eta$ that of the ``disconnected'' pair correlation function at criticality.) 
Their dependence on the spatial dimension $d$ is shown in Fig.~\ref{fig_eta}. 
Above $d_{DR}\simeq 5.1$, we find that $\eta$ and $\bar \eta$
rigorously coincide, $\bar \eta=\eta$, which is the signature of
dimensional reduction. (Their value is moreover equal to that found in
the same order of the derivative expansion for the pure Ising model in
dimension $d-2$.) Below $d_{DR}$, the two anomalous dimensions
bifurcate and $\bar \eta \neq \eta$.  Note that the predicted values
satisfy the required bounds, \textit{i.e.}, $2 \eta \geq \bar \eta
\geq \eta$, $\eta \geq (4-d)/2$, $\bar \eta \geq 4-d$. 

We give in Table~\ref{tab_results_eta_etab} our estimates for $\eta$ and $\bar \eta$ in $d=3$ and
$d=4$, where a comparison is possible with numerical determinations
from computer studies (Monte Carlo calculations and $T=0$ ground-state determinations).
\begin{table}[hb]
\begin{tabular}{|l|l|l|l|}
\hline
$d$&$\eta$ (this work)&$\eta$ (ground state)&$\eta$ (Monte Carlo)\\
\hline
3&0.57(2)&$0.51^{a}$&0.5-0.6$^{f,g,h}$\\
4&0.24(1)&0.18$^{b}$,0.20$^{c}$?&....\\
\hline
\end{tabular}

\begin{tabular}{|l|l|l|l|}
\hline
$d$&$\bar\eta$ (this work)&$\bar\eta$ (ground state)&$\bar\eta$ (Monte Carlo)\\
\hline
3&1.08(2)&1.02$^{a}$,1.06$^{d}$,1.1$^{e}$&1.0-1.04$^{f,g,h}$\\
4&0.40(?)&0.37$^{b}$,0.38$^{c}$&...\\
\hline
\end{tabular}
\begin{tabular}{|l|l|l|l|}
\hline
$d$&$\nu$ (this work)&$\nu$ (ground state)&$\nu$ (Monte Carlo)\\
\hline
3&1.08(2)&$1.0^{e}$,$1.1^{i}$,$1.19^{d}$&$1.0^{m}$,$1.1^{g}$,$1.31^{n}$\\
&&,$1.22^{j}$,$1.25^{k}$,$1.36^{l}$&\\
&&$1.37^{a}$&\\
4&0.81(3)&$0.78^{b}$,$0.82^{c}$&\\
\hline
\end{tabular}
\caption{\label{tab_results_eta_etab}
Comparison between the anomalous dimensions obtained in the present work and in computer studies (a: [\onlinecite{middleton-fisher02}]; b: [\onlinecite{hartmann02}]; c: [\onlinecite{middleton02}]; d: [\onlinecite{hartmann99}]; e: [\onlinecite{ogielski86}]; f: [\onlinecite{rieger93}]; g: [\onlinecite{rieger95}]; h: [\onlinecite{fytas11}]; i: [\onlinecite{dukovski03}]; j: [\onlinecite{angles97}]; k: [\onlinecite{wu06}]; l: [\onlinecite{hartmann01}]; m: [\onlinecite{nowak98}]; n: [\onlinecite{malakis06}]). The error bars on the results from simulations and ground-state computations can be found in the original papers.The values of the temperature exponent $\theta=2+\eta - \bar\eta$ that we obtain are $1.49$ in $d=3$ and $1.84$ in $d=4$, in excellent agreement with the other studies.}
\end{table}
One can see that the agreement is good or very good. (Real-space RG studies in $d=3$\cite{dayan93,newman93,cao93,falikov95,efrat03} provide results in the same range, with an anomalous dimension $\eta$ between $0.51$ to $0.56$ and an exponent $\theta$ between $1.50$ and $1.56$.) As is well known
from the study of simpler models such as the pure $O(N)$ model, the
way to further improve the accuracy of the exponents would be to
consider higher orders of the approximation scheme presented in
section IV-A.

We have not been able to perform our NP-FRG calculation down to the lower critical dimension $d=2$, where the values $\eta=1$ and $\bar \eta=2$ are exactly known. The numerical resolution of the flow equations become extremely arduous in low dimension where the anomalous dimensions become large and approach their lower bound, $(4-d)/2$ and $4-d$ respectively. In addition, one also encounters numerical difficulties as one approaches the critical dimension $d_{DR}$ from below. Indeed, as mentioned in section IV-D, it is  anticipated that the limit $d\rightarrow d_{DR}^-$, $y\rightarrow 0$ is nonuniform with a boundary layer in $y^2/(d_{DR}-d)$. More work will be needed to solve this boundary-layer problem numerically.

Fig.~\ref{fig_eta} also provides evidence that the claim according to which the two exponents $\eta$ and $\bar \eta$ are related by a fixed ratio, $\bar{\eta}=2 \eta$,\cite{schwartz85b,soffer85,schwartz91,gofman93} cannot be right in general. It is true that the relation $\bar{\eta}=2 \eta$ is exact in $d=2$ (and according to a phenomenological RG,\cite{bray85} also at first order in $d=2+\epsilon$) and is very closely obeyed by the numerical estimates in $d=3$ and $d=4$ (see Table~\ref{tab_results_eta_etab}). However, the latter type of ``numerical evidence'' can always be challenged, and it is actually impossible to reach a definite conclusion on the sole basis of numerical results in selected dimensions ($d=3,4,5$). On the other hand, the overall and continuous dependence on spatial dimension that we provide through the NP-FRG leads to a different and firmer type of answer: since one goes from $\bar \eta=\eta$ for $d\geq d_{DR}$ to $\bar \eta>\eta$ above, the relation  $\bar{\eta}=2 \eta$ cannot be always valid.

Finally, we also display our results in $d=3$ and $d=4$ for the correlation length exponent $\nu$, which is associated with the relevant direction around the critical fixed point, in Table I. The agreement with the available data from computer studies is excellent in $d=4$ and fair in $d=3$. (Note that in the real-space RG studies in $d=3$\cite{dayan93,newman93,cao93,falikov95,efrat03}, the exponent $\nu$ is in general too large, with values ranging from $1.39$ to $2.25$.) All the other critical exponents are obtained in the NP-FRG through the expected relations:  $\gamma=(2-\eta)\nu$, $\bar \gamma=(4-\bar \eta)\nu$, $\delta=(d+4-\bar \eta)/(d-4+\bar \eta)$, etc.\cite{footnote7}

The validity of the theoretical description is clearly confirmed by both the overall consistency and  the quantitative accuracy of the predictions in $d=3$ and $d=4$.

\subsection{Role of the auxiliary temperature and correction to ``Grassmannian ultralocality''}

To assess the effect of a nonzero auxiliary temperature and of a deviation from  ``Grassmannian ultralocality'', we have considered the simplest ``non-ultralocal'' contribution to the cumulants introduced in section VII-B of  paper III. Combined with the above truncation of the ``ultralocal'' contributions of effective average action [Eqs.~(\ref{eq_ansatz1},\ref{eq_ansatz2})], this amounts to expressing the first cumulant as (see Eq. (136) of paper III)
\begin{equation}
\begin{aligned}
\label{eq_cumulant_1_nonultralocal_correction}          
&\mathsf \Gamma_{k1}[\Phi_{1}] =  \int_{\underline{\theta}_1}
\bigg (\Gamma_{k1}[ \Phi_1(\underline{\theta}_1)]+\frac{1}{2\beta}(1+ \beta \bar \theta_1 \theta_1)\times \\& \int_x Y_{k}(\Phi_1(x,\underline{\theta}_1))\partial_{ \theta_1} \Phi_1(x,\underline{\theta}_1)\partial_{\bar \theta_1} \Phi_1(x,\underline{\theta}_1) \bigg ),
\end{aligned}
\end{equation}
with $\Gamma_{k1}[\phi]$ given in Eq.~(\ref{eq_ansatz1}), whereas the second-order cumulant is taken as purely ``ultralocal'' and given in Eq.~(\ref{eq_ansatz2}). The FRG flow equation for the function $Y_k(\phi)$ is directly obtained from Eq.~(139) of paper III. In a dimensionless form, it reads 
\begin{equation}
\begin{split}
\label{eq_flow_ydimensionless}
&\partial_t y_k(\varphi)=(2\eta_k-\etab_k)y_k(\varphi)+\frac{d-4+\etab_k}2\varphi y_k'(\varphi)+ \\&v_d\Big(2
y_k''(\varphi)-\frac{y_k'(\varphi)^2}{y_k(\varphi)}\Big)\Big\{(d-2) l^{(d-2)}_1(\varphi)+2[\delta_{k,0}(\varphi)\\&-z_k(\varphi)] l^{(d)}_2(\varphi)+(\bar \eta_k-\eta_k)j^{(1,d)}_2(\varphi)\Big\}+ O(T_k),
  \end{split}
\end{equation}
where the threshold functions are defined in Appendix~\ref{appendixB}. In deriving the above expression from Eq.~(139) of the preceding paper, we have used the dimensionless quantity $y_k$ defined through $Y_k(\phi)=(Z_k k^2/T_k)y_k(\varphi)$ and the fact that as $k \rightarrow 0$,
\begin{equation}
\begin{aligned}
\label{eq_hatQP_approx}          
\widehat Q_k(q^2,\phi)-\widehat P_k(q^2,\phi)&=\frac{1}{\widehat P_k(q^2,\phi)^{-1}+Y_k(\phi)} \\& \sim  \frac{1}{Y_k(\phi)} \sim  \frac{k^{2\eta-\bar\eta}}{y_k(\varphi)}.
\end{aligned}
\end{equation}
The above equation is valid provided that $\widehat P_k^{-1}$ which asymptotically goes as $k^{2-\eta}$ is subdominant with respect to $Y_k$. As discuss just below, this is indeed true, although $y_k$ goes to zero at the fixed point.

The flow of the ``ultralocal'' quantities in the limit $\beta^{-1}=0$  has already been solved and we can introduce the corresponding solution in Eq.~(\ref{eq_flow_ydimensionless}). We have solved the latter equation and found that $y_k(\varphi)$ goes to zero as one approaches the fixed point, in such a way that the dimensionful quantity $Y_k(\varphi)$ goes to a constant $Y_*$: this is illustrated in Fig. 10 for $d=3$.

\begin{figure}[htbp]
  \centering
 \includegraphics[width=\linewidth]{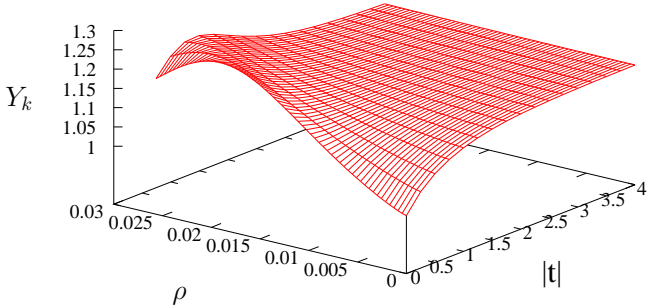}
  \caption{Flow of the (dimensionful) function $Y_k$ of the
    dimensionless field with $\rho= \varphi^2/2$. After a transient regime, the function tends to a constant.}
\end{figure}

In addition, we can investigate the effect of the ``non-ultralocal'' contribution to the flow of the ``ultralocal'' quantities when the auxiliary temperature $\beta^{-1}$ is different from zero. This can be done by considering the  ERGE's for the ``ultralocal'' components of the first and the second cumulants given in  Eqs.~(134) and (135) of paper III. From the above results (in particular, the fact that $Y_k$ goes to a finite constant when $k \rightarrow 0$), we know that $\widehat Q_k(q^2,\phi)$ is asymptotically equal to $\widehat P_k(q^2,\phi)$: see Eq.~(\ref{eq_hatQP_approx}) and discussion below. The flow of the dimensionless first cumulant therefore follows Eq.~(151)  of paper III and the contribution of the ``non-ultralocal'' piece goes as $T_k$ times a well-behaved function(al), with $T_k\sim k^{\theta}/\beta$ and $\theta=2+\eta-\bar \eta >0$.

The flow of the second cumulant is also of the form of Eq.~(151) in the preceding paper, but the dimensionless beta function due to the ``non-ultralocal'' piece is now potentially singular when a cusp appears in the limit of zero (auxiliary) temperature  of the second cumulant. More specifically, one finds that
\begin{equation}
 \label{eq_v_isingBL}
\begin{split}
&\partial_t \delta_k(\varphi_1,\varphi_2)=\beta_{\delta}^{\beta=\infty}(\varphi_1,\varphi_2)+\\&\frac{ T_k}{2}\widetilde \partial_t\int_{\hat q}\Big\{\partial_{\varphi_1} \left [\frac{\delta_k^{(10)}(\varphi_1,\varphi_2)}{z_k(\varphi_1)\hat q^2+s(\hat q^2) +u''_k(\varphi_1)}\right ]+perm(12) \Big\},
\end{split}
\end{equation}
where $\beta_{\delta}^{\beta=\infty}(\varphi_1,\varphi_2)$ is the zero (auxiliary) temperature beta functional obtained by taking derivatives with respect to $\varphi_1$ and $\varphi_2$ of the right-hand side of Eq.~(\ref{eq_v_ising}) and we have omitted  subdominant terms coming from the difference between $\widehat Q_k$ and $\widehat P_k$ (see above).

We can now follow the derivation done for the effect of the (bath) temperature in the formalism without superfields (see paper II\cite{tissier08}). After changing variables from $\varphi_1, \varphi_2$ to $x=(\varphi_1+ \varphi_2)/2$ and $y=(\varphi_1- \varphi_2)/2$, one obtains the solution for $\delta_k(x,y)$ in the close vicinity of the fixed point, when both $T_k$ and $y$ go to zero, in the form of a boundary layer
\begin{equation}
 \label{eq_boundary_layer2}
\begin{split}
&\delta_k(x,y)=\\&\delta_{*,0}(x)+ T_k \left [A(x)-B(x)\sqrt{1+C(x)^2\left (\frac{y}{T_k}\right)^2}\right ] + O(T_k^2),
\end{split}
\end{equation}
where $A(x)$, $B(x)$ , $C(x)$ are well-behaved functions of $x$ and $A(x)\vert C(x)\vert $ is equal to the absolute value of the coefficient of the cusp in the zero-temperature fixed-point function $\delta_{*}(x,y)$.

The connection between this boundary-layer phenomenon and the presence of rare low-energy excitations above the ground-state known as ``droplets''\cite{bray84,fisher-huse88} is the same as that already discussed in our previous work\cite{tissier08} (see also Refs.~[\onlinecite{BLbalents04},\onlinecite{BLledoussal10}]) and the discussion is not repeated here.

\section{Conclusion}
\label{sec:conclusion}

In this paper and the preceding one \cite{tissier11b} we have extended our NP-FRG approach of disordered systems, which was initiated in the previous articles of this series.\cite{tarjus08,tissier08} The objective was to discuss the property of dimensional reduction and its breakdown in the RFIM from the Parisi-Sourlas\cite{parisi79} perspective of an underlying supersymmetry and its breaking. We have reformulated the NP-FRG in a superfield formalism and, to do so, we have proposed a solution for properly selecting the ground state among the many metastable states that are present at zero temperature. Through the introduction of an appropriate regulator and a supersymmetry-compatible nonperturbative approximation, we have been able to follow the supersymmetry (more precisely, the superrotational invariance) and its spontaneous breaking along the RG flow.

Despite the fact that the effective hamiltonian (or microscopic action) in the presence of a random field has numerous minima in the region of interest near the critical point, dimensional reduction need not be systematically broken. By implementing a NP-FRG flow for the cumulants of the renormalized disorder, our work shows that there is a finite range of dimension below the upper critical dimension, $d_{DR} \leq d \leq d_{uc}=6$, for which this multiplicity of minima has no effect on the long-distance properties of the model. More precisely, the scaling behavior around the critical point conforms to the dimensional-reduction predictions. The associated fixed point is characterized by a nonanalytic dependence of the effective action in the dimensionless fields, at odds with a naive description based on perturbation theory, but the nonanalyticity is too weak to alter the spectrum of critical exponents. It is only by going below a critical dimension $d_{DR}$, which we numerically find around $5$, that spontaneous breaking of the supersymmetry (superrotational invariance) takes place. The robustness of our theoretical description, which explains the pending puzzles concerning the critical behavior of the RFIM, is supported by the good agreement obtained between the NP-FRG predictions
for the critical exponents and the available values from computer studies.

Finally, one may wonder whether the formalism developed here can be useful in a different context. An obvious extension is a study of the long-distance physics of self-avoiding branched polymers and the associated property of reduction to the Yang-Lee edge singularity problem in two fewer dimensions.\cite{parisi81} Dimensional reduction is known to hold in this case,\cite{imbrie03} and this provides a benchmark model to test the ability of our nonperturbative approximation scheme to reproduce this property. Another extension concerns the hysteresis behavior and out-of-equilibrium phase transitions of the RFIM at zero temperature when driven by a slow change of the external magnetic field.\cite{sethna05,vives04,liu09} 
Work in those directions is under way.

More challenging, and far more speculative too, is the question known
as the ``Gribov ambiguity'' in the nonperturbative quantization of
nonabelian gauge field
theories.\cite{Gribov77,zinnjustin89,esposito04} The standard
Faddeev-Popov gauge-fixing procedure aims at restricting the
functional integral over the gauge field to nonequivalent gauge
configurations (\textit{i.e}, configurations that are not obtainable
one from another by a gauge transformation). Unfortunately, the
gauge-fixing conditions that respect Lorentz invariance and internal
(color) symmetry, such as the Landau gauge condition, are inconsistent
because they admit solutions that are equivalent up to a gauge
transformation (``Gribov copies''). There has been a strong
theoretical effort to overcome this problem. This ``Gribov
ambiguity'' does not affect the perturbative regime, so that calculations at high energy 
can be performed in the standard
Faddeev-Popov approach. However, very little is known on its influence
in the nonperturbative regime. The Gribov-Zwanziger
model~\cite{Gribov77,Zwanziger89} enables one to reduce the
number of copies taken into account in the functional, but not to a single one, and, up to now, there is no
unambiguous gauge-fixing procedure. The strong connections between the
Faddeev-Popov and the Parisi-Sourlas formalisms, and their common
failure when the relevant field equation has a multiplicity of
solutions, have been underlined.\cite{zinnjustin89} The relevance of
the tools and concepts developed in the present and the preceding
papers to the problem of the Gribov ambiguity has yet to be
investigated.

\appendix

\section{Stability of the nonanalytic behavior along the RG flow}
\label{appendixA}

To illustrate that a linear cusp in $\Gamma_{k2;x_1,x_2}^{(11)}(\phi_1,\phi_2)$ does not generate a supercusp and may break dimensional reduction, we consider the ERGE for the Fourier transform of the 2-point proper vertex $\Gamma_{k2}^{(11)}(q^2;\phi_1,\phi_2)$ (for uniform fields) in the limit $\phi_2\rightarrow \phi_1$. We switch to the variables $\phi=(\phi_1+\phi_2)/2$ and $Y=(\phi_1-\phi_2)/2$ and we study a situation in which a cusp is present:
\begin{equation}
\label{eq_cusp_stab}
\Gamma_{k2}^{(11)}(q^2;\phi,Y) = \Gamma_{k2;0}^{(11)}(q^2;\phi) + \Gamma_{k2;\alpha}^{(11)}(q^2;\phi) |Y|^{\alpha} + \cdots,
\end{equation}
where $Y \rightarrow 0$ and $\alpha < 2$ (recall that $\Gamma_{k2}^{(11)}$ is even in $\phi$ and $Y$ separately). With the  assumption concerning the behavior of the proper vertices detailed in Sec.~III-B, we find that the flow of $\Gamma_{k2}^{(11)}$, where for simplicity we take $q^2=0$, can be written in the $Y \rightarrow 0$ limit:
\begin{equation}
\begin{aligned}
\label{eq_flow_cusp_stab}
\partial_t &\Gamma_{k2}^{(11)}(q^2=0;\phi,Y) \simeq regular -\frac{\alpha^2}{4} |Y|^{2(\alpha-1)}\times \\& \tilde \partial_t \int_{q'} \widehat P_k(q'^2;\phi)^2 \Gamma_{k2;\alpha}^{(11)}(q'^2;\phi)^2 - \tilde \partial_t \int_{q'}\widehat P_k(q'^2;\phi)\\& \times \big[\partial_{Y}\Gamma_{k3;-q'q'0}^{(111)}(\phi+Y,\phi+Y,\phi_3)|_{\phi_3=\phi-Y}) \\&- \partial_{Y}\Gamma_{k3;-q'q'0}^{(111)}(\phi-Y,\phi-Y,\phi_3)|_{\phi_3=\phi+Y}) \big],
\end{aligned}
\end{equation}
where $regular$ denotes the terms that could be obtained by directly considering the flow of $\Gamma_{k2}^{(11)}(q^2=0;\phi,Y=0)$ and by assuming a regular behavior of its beta-functional when $Y \rightarrow 0$. The contribution due to the derivatives of the third cumulant $\Gamma_{k3}^{(111)}$ is an even function of $Y$; the $Y=0$ term contributes to the regular component of the beta-functional and the next term is a $O(|Y|^{\alpha'}$) with $\alpha' > 2$.

The above equation indicates that a ``supercusp'' with $\alpha <1$ leads to an ill-defined flow for $\Gamma_{k2}^{(11)}(q^2=0;\phi,Y=0)$. For $\alpha \geq1$ (cusp) one should also study the ERGE for the third cumulant to determine the leading nonanalyticity (\textit{i.e.}, $\alpha$') and therefore study the stability of the cusp under further RG flow. In any case, Eq.~(\ref{eq_flow_cusp_stab}) shows that a cusp weaker than linear, with $1< \alpha <2$, has no effect on the flow of $\Gamma_{k2}^{(11)}(q^2=0;\phi,Y=0)$ and, since $2 \eta_k - \bar \eta_k$ is obtained from the latter, is not expected to modify dimensional reduction. A linear cusp ($\alpha =1$) on the other hand  provides a possible mechanism for the failure of dimensional reduction while not \textit{a priori} generating stronger nonanalyticities. In order to actually lead to breakdown of dimensional reduction, the linear cusp must of course remain in the renormalized second cumulant up to the appropriate fixed point.

\section{Dimensionless threshold functions}
\label{appendixB}

The various dimensionless threshold functions correspond to the various $1$-loop integrals involving the infrared cutoff functions and the propagators (after account of the scaling dimensions). More specifically, they are defined as follows:\cite{berges02,tetradis94,delamotte03}
\begin{equation}
\begin{split}
  \label{eq_l_2}
 l&_{q_1,q_2}^{(d)}(w_1,w_2;z_1,z_2)=\\&- \frac 12  \int_0^\infty 
  dy\,y^{d/2-1}\tilde \partial_t \left\lbrace  \frac{1}{(p_1(y)+w_1)^{q_1}(p_2(y)+w_2)^{q_2}}\right\rbrace ,
\end{split}
\end{equation}
\begin{equation}
  \begin{split}
  \label{eq_threshold_m}
m&_{q_1,q_2}^{(d)}(w_1,w_2;z_1,z_2)=\\&- \frac 12  \int_0^\infty 
dy\,y^{d/2-1}\tilde \partial_t \left\lbrace  \frac{y\left(\partial_y p_1(y) \right) ^2}{(p_1(y)+w_1)^{q_1}(p_2(y)+w_2)^{q_2}}\right\rbrace,
      \end{split}
\end{equation}
\begin{equation}
  \begin{split}
  \label{eq_threshold_n}
 n&_{q_1,q_2}^{(d)}(w_1,w_2;z_1,z_2)=\\&- \frac 12  \int_0^\infty  dy\,y^{d/2-1}
 \tilde \partial_t \left\lbrace \frac{y \partial_y p_1(y)}{(p_1(y)+w_1)^{q_1}(p_2(y)+w_2)^{q_2}}\right\rbrace,
 \end{split}
\end{equation}
with $p_a(y)=y(z_a+r(y)), a=1,2$, $y=q^2/k^2$, and $\tilde \partial_t$
acts only on the dimensionless function $r(y)=s(y)/y=\widehat
R_k(yk^2)/Z_kyk^2$ that is contained in the
$p_a(y)$'s\cite{wetterich93,berges02,delamotte03}(by definition, $\partial_t r(y)=-[\eta r(y)+2 y r'(y)]$). Note that, in
addition to the dependence on $w\equiv u_k''(\phi)$ and $z\equiv
z_k(\phi)$, the threshold functions explicitly depend on the scale $k$
via the running anomalous dimension $\eta_k$. (Note also that the
functions $m^{(d)}$ and $n^{(d)}$ are not symmetric in the exchange of
the indices $1$ and $2$.) The generic properties of these
dimensionless threshold functions are discussed in detail in
Ref.~[\onlinecite{berges02,tetradis94,delamotte03}].

In the present study, it is also necessary to introduce additional threshold
functions, which accounts for the fact that the cutoff function
$\widetilde R$ has an anomalous scaling compared to $\widehat R$
when dimensional reduction is broken. These threshold functions then always appear
multiplied by $(\eta-\etab)$ in the flow equations. We define:
\begin{equation}
  \begin{split}
  \label{eq_threshold_j}
j&_{q_1,q_2}^{(a,d)}(w_1,w_2;z_1,z_2)=\\&-   \int_0^\infty 
dy\,y^{d/2-1}  \frac{(r(y)+y r'(y))^a}{(p_1(y)+w_1)^{q_1}(p_2(y)+w_2)^{q_2}}
      \end{split}
\end{equation}
and
\begin{equation}
  \begin{split}
  \label{eq_threshold_h}
h_{q}^{(d)}(w;z)&=\\-   \int_0^\infty 
dy&\,y^{d/2}  \frac{r(y)+y r'(y)}{(p(y)+w)^{q}}\Big\{\frac{2z+r(y)+y r'(y)}2- \\&\frac{qzy}{p(y)+w}(z+r(y)+yr'(y))\Big\}.
      \end{split}
\end{equation}

We summarize here a number of relations that are useful in the developments of Sec. IV. One has  $l_q^{(d)}(w;z)= l_{q_1,q_2}^{(d)}(w,w;z,z)$, $m_q^{(d)}(w;z)= m_{q_1,q_2}^{(d)}(w,w;z,z)$, $n_q^{(d)}(w;z)=n_{q_1,q_2}^{(d)}(w,w;z,z)$, $j_q^{(a,d)}(w;z)=j_{q_1,q_2}^{(a,d)}(w,w;z,z)$, with $q_1 + q_2 =q$. In addition, the functions $n_q^{(d)}$'s and $l_q^{(d)}$'s are related by
\begin{equation}
n_{q}^{(d)}(w;z)= \frac{d}{2(q-1)}l_{q-1}^{(d)}(w;z).
\end{equation}

From the above definitions, one also straightforwardly finds that
\begin{subequations}
\begin{equation}
\partial_{w_1}l_{q_1,q_2}^{(d)}(1,2)= -q_1  l_{q_1+1,q_2}^{(d)}(1,2),
\end{equation}

\begin{equation}
\partial_{z_1}l_{q_1,q_2}^{(d)}(1,2)= -q_1  l_{q_1+1,q_2}^{(d+2)}(1,2),
\end{equation}
\end{subequations}
and
\begin{subequations}
\begin{equation}
\partial_{w_1}m_{q_1,q_2}^{(d)}(1,2)= -q_1  m_{q_1+1,q_2}^{(d)}(1,2),
\end{equation}
\begin{equation}
\partial_{z_1}m_{q_1,q_2}^{(d)}(1,2)= -q_1  m_{q_1+1,q_2}^{(d+2)}(1,2) + \frac{d}{q_1-1}l_{q_1-1,q_2}^{(d+2)}(1,2),
\end{equation}
\begin{equation}
\partial_{z_2}m_{q_1,q_2}^{(d)}(1,2)= -q_2  m_{q_1,q_2+1}^{(d+2)}(1,2).
\end{equation}
\end{subequations}
Similar relations hold for the functions $j_{q_1,q_2}^{(a,d)}$ and $h_q^{(d)}$. From these  expressions one can obtain the derivatives with respect to the field arguments $\varphi_a$ by using that $w\equiv u_k''(\phi)$ and $z\equiv z_k(\phi)$.

\section{Some technical aspects of the numerical resolution}
\label{appendix C}

\subsection{Optimization of the IR cutoff function}

\begin{figure}[htbp]
\includegraphics[width=.9\linewidth]{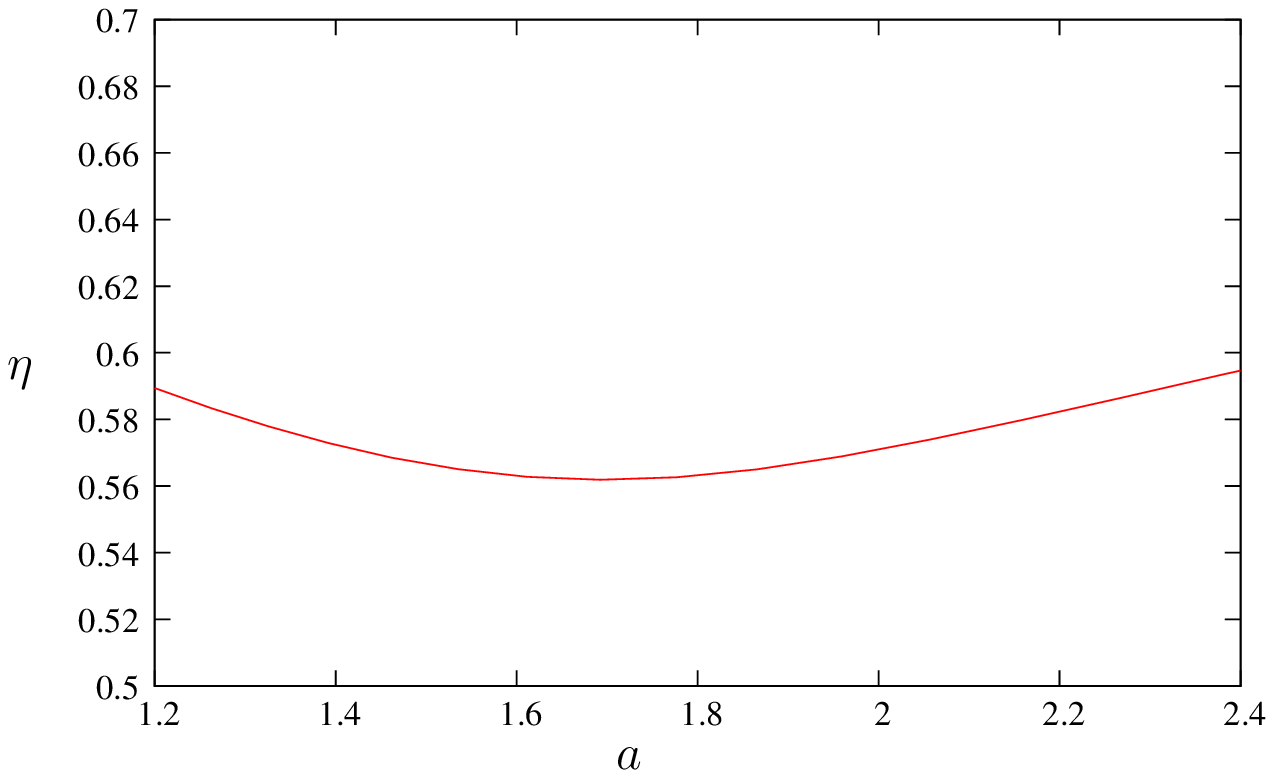}
\includegraphics[width=.9\linewidth]{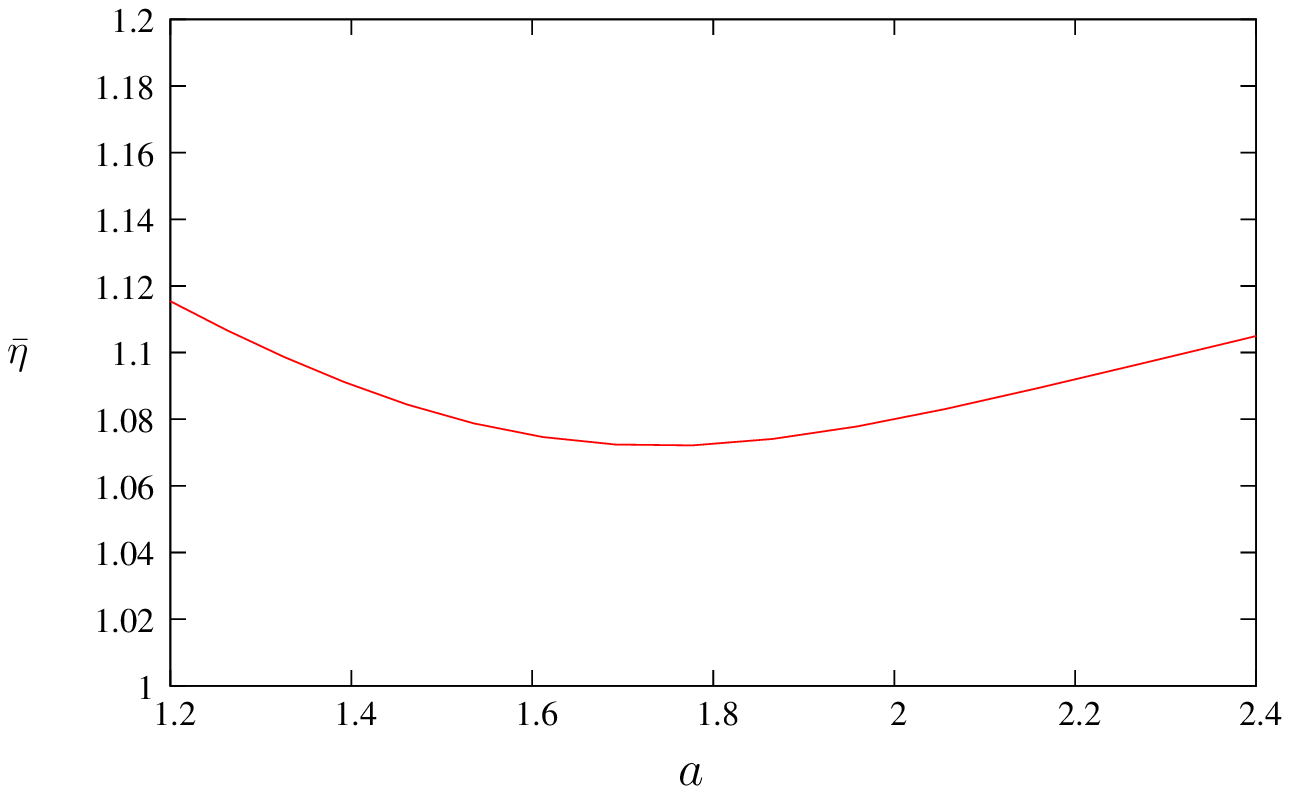}
\includegraphics[width=.9\linewidth]{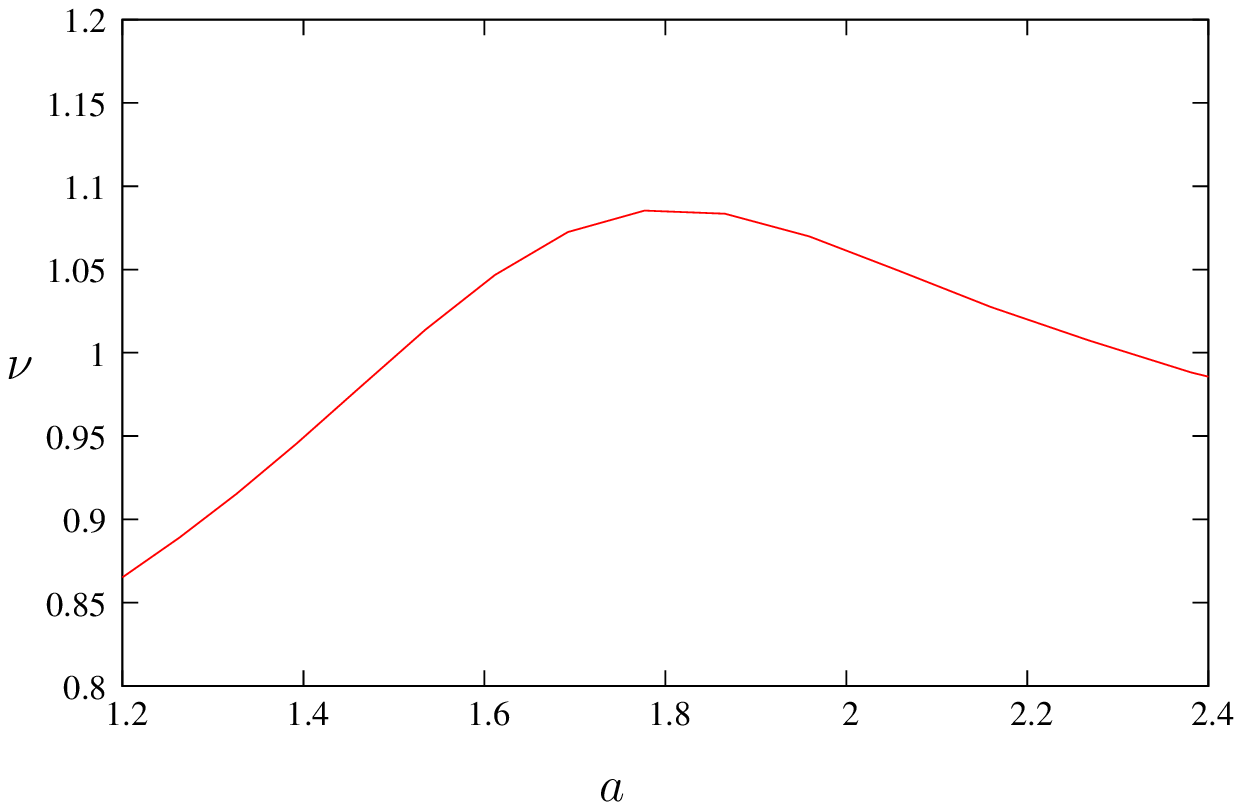}
\caption{\label{fig_etapms} Dependence of the anomalous dimensions
  $\eta$ (upper panel), $\bar \eta$ (middle panel) and of the critical
  exponent $\nu$ (lower panel) on the parameter $a$ of the IR cutoff
  function in Eq.~(\ref{eq_chosen_r}). The ``principle of minimal
  sensitivity'' leads to a determination of the anomalous exponents
  $\eta=0.565$, $\bar\eta=1.075$, $\nu=1.08$ (at the minima). The
  variations of the anomalous dimensions are very small over a wide
  range of $a$ and enables us to evaluate the precision to
  $\Delta\eta=0.03$ and $\Delta\bar\eta=0.04$. $\nu$ is more sensitive
to the regulator parameters and we estimate $\Delta\nu=0.2$}
\end{figure}
In the exact formulation of the NP-FRG, the results of the flow
equations are independent of the IR regulator, which is only an
intermediate means that does not affect the long-distance
physics. Once approximations are introduced, a residual dependence on
the choice of regulator however remains. (A similar situation occurs
in perturbation theory where a residual dependence on the parameters
of the Borel resummation procedure is observed.) One can in some sense
optimize the choice of IR cutoff function by demanding that the
output, say the critical exponents, satisfies a property of ``minimal
sensitivity'' such that by varying the characteristics of the cutoff function
around the optimum, a minimal variation of the exponents results. This
guarantees the stability of the results. We apply this procedure to
the type of function given in Eq.~(\ref{eq_chosen_r}) where the
parameters $a$, $b$, and $c$ can be varied.

We illustrate in Fig.~\ref{fig_etapms} the dependence of the anomalous dimensions in $d=3$ on the parameter $a$ when $b$ and $c$ are kept fixed at the values $0.81$ and $0.14$ respectively (these latter values were determined by preliminary variational studies to find minimal sensitivity). One can see that there is a large domain of values in which $\eta$ and $\bar \eta$ vary little (say by less than $5 \%$). We use the ``principle of minimal sensitivity'' to determine the best estimate of the critical exponents [note that the minima of $\eta(a)$ and $\bar\eta(a)$ happen at very close values of $a$]. We also estimate the error bars  from the observation that when $a$ varies in the range [1.5,3], the maximum variation of $\eta$ is of 0.03 and that of $\bar\eta$ is of 0.04. These are the values that we have reported in Sec.~\ref{sec:results}.

\subsection{Regularization of the numerical instabilities}

During the numerical integration of the flow, we encountered 
numerical instabilities in physically unimportant regions (typically,
at large fields). The origin of these instabilities can be understood
as follows. In the region of large fields, the threshold functions
rapidly decrease to zero. In this limit, the flow of a function is
given by the dimensional part, {\it i.e.} the flow obtained by setting $v_d\equiv 0$ [\textit{e.g.} $\partial_t
u_k' (\varphi)\sim-\frac{1}{2}(d-2\eta_k+\bar\eta_k)u_k'(\varphi)+
\frac{1}{2}(d-4+\bar\eta_k) \varphi\; u_k''(\varphi)$]. 
This flow no longer depends on the second derivative of the associated
function with respect to the field. However, this diffusive-like term [$u_k'''(\varphi)$ in our previous example] is very
important to stabilize the numerical integration. In practice, when
necessary, we have therefore added by hand a small diffusive contribution to the
flow equations and checked that the numerical results are stable under
varying the strength of these extra terms.


\begin{thebibliography}{15}
\expandafter\ifx\csname natexlab\endcsname\relax\def\natexlab#1{#1}\fi
\expandafter\ifx\csname bibnamefont\endcsname\relax
  \def\bibnamefont#1{#1}\fi
\expandafter\ifx\csname bibfnamefont\endcsname\relax
  \def\bibfnamefont#1{#1}\fi
\expandafter\ifx\csname citenamefont\endcsname\relax
  \def\citenamefont#1{#1}\fi
\expandafter\ifx\csname url\endcsname\relax
  \def\url#1{\texttt{#1}}\fi
\expandafter\ifx\csname urlprefix\endcsname\relax\def\urlprefix{URL }\fi
\providecommand{\bibinfo}[2]{#2}
\providecommand{\eprint}[2][]{\url{#2}}

\bibitem[{\citenamefont{Tarjus and Tissier}(2008)}]{tarjus08}
\bibinfo{author}{\bibfnamefont{G.}~\bibnamefont{Tarjus}} \bibnamefont{and}
  \bibinfo{author}{\bibfnamefont{M.}~\bibnamefont{Tissier}},
  \bibinfo{journal}{Phys. Rev. B} \textbf{\bibinfo{volume}{78}},
  \bibinfo{pages}{024203} (\bibinfo{year}{2008}).

\bibitem[{\citenamefont{Tissier and Tarjus}(2008)}]{tissier08}
\bibinfo{author}{\bibfnamefont{M.}~\bibnamefont{Tissier}} \bibnamefont{and}
  \bibinfo{author}{\bibfnamefont{G.}~\bibnamefont{Tarjus}},
  \bibinfo{journal}{Phys. Rev. B} \textbf{\bibinfo{volume}{78}},
  \bibinfo{pages}{024204} (\bibinfo{year}{2008}).

\bibitem{footnote00}
The physics associated with ``Griffiths phases" is however unlikely to be captured by such an approach. In this case the rare regions that are involved are exponentially suppressed as their size increases and produce very weak (essential) singularities in the thermodynamics. Such a behavior does not show up easily once the average over disorder has been performed, as done in the cumulants. Droplets on the other hand are ``power-law rare'' regions in their size and have a clear signature in the functional dependence of the cumulants of the renormalized disorder (see \textit{e.g.} the detailed work by Balents and Ledoussal on the random manifold model.\cite{BLbalents04,BLledoussal10}) The same is true for ``avalanches", which are collective discontinuous events occurring in \textit{typical} samples at zero temperature for exceptional values of the applied source.

\bibitem{BLbalents04}
L. Balents and P. Ledoussal, Europhysics Lett. \textbf{65}, 685 (2004); Ann. Phys. \textbf{315}, 213 (2005).

\bibitem{BLledoussal10}
P. Ledoussal, Ann. Phys. \textbf{325}, 49 (2010). 

\bibitem{fisher86a}
D. S. Fisher, Phys. Rev. Lett. \textbf{56}, 1964 (1986).

\bibitem{balents-fisher93}
L. Balents and D. S. Fisher, Phys. Rev. B \textbf{48}, 5949 (1993).

\bibitem{balents96}
L. Balents, J. P. Bouchaud, and M. M\'ezard, J. Phys. I \textbf{6}, 1007 (1996).

\bibitem{BLchauve00}
P. Chauve, T. Giamarchi, and P. Ledoussal, Phys. Rev. B \textbf{62}, 6241 (2000).

\bibitem{CUSPledoussal}
P. Le Doussal, K. J. Wiese, and P. Chauve, Phys. Rev. B \textbf{66}, 174201 (2002); Phys. Rev. E \textbf{69}, 026112 (2004).

\bibitem{CUSPledoussal09}
P. Le Doussal and K. J. Wiese, Phys. Rev. E \textbf{79}, 051106 (2009). 

\bibitem{villain84}
J. Villain, Phys. Rev. Lett. \textbf{52}, 1543 (1984).

\bibitem{fisher86}
D. S. Fisher, Phys. Rev. Lett. \textbf{56}, 416 (1986).

\bibitem[{\citenamefont{Nattermann}(1998)}]{nattermann98}
\bibinfo{author}{\bibfnamefont{T.}~\bibnamefont{Nattermann}},
  \emph{\bibinfo{title}{Spin glasses and random fields}}
  (\bibinfo{publisher}{World scientific, Singapore}, \bibinfo{year}{1998}), pp.
  \bibinfo{pages}{277, and references therein}.

\bibitem[{\citenamefont{Parisi and Sourlas}(1979)}]{parisi79}
\bibinfo{author}{\bibfnamefont{G.}~\bibnamefont{Parisi}} \bibnamefont{and}
  \bibinfo{author}{\bibfnamefont{N.}~\bibnamefont{Sourlas}},
  \bibinfo{journal}{Phys. Rev. Lett.} \textbf{\bibinfo{volume}{43}},
  \bibinfo{pages}{744} (\bibinfo{year}{1979}).

\bibitem[{\citenamefont{Imbrie}(1984)}]{imbrie84}
\bibinfo{author}{\bibfnamefont{J.~Z.} \bibnamefont{Imbrie}},
\bibinfo{journal}{Phys. Rev. Lett.} \textbf{\bibinfo{volume}{53}},
 \bibinfo{pages}{1747} (\bibinfo{year}{1984}).

\bibitem[{\citenamefont{Bricmont and Kupianen}(1987)}]{bricmont87}
\bibinfo{author}{\bibfnamefont{J.}~\bibnamefont{Bricmont}} \bibnamefont{and}
 \bibinfo{author}{\bibfnamefont{A.}~\bibnamefont{Kupianen}},
  \bibinfo{journal}{Phys. Rev. Lett.} \textbf{\bibinfo{volume}{59}},
  \bibinfo{pages}{1829} (\bibinfo{year}{1987}).

\bibitem[{\citenamefont{Parisi}(1984)}]{parisi84b}
\bibinfo{author}{\bibfnamefont{G.}~\bibnamefont{Parisi}}, in
  \emph{\bibinfo{booktitle}{Proceedings of Les Houches 1982, Session XXXIX}},
  edited by \bibinfo{editor}{\bibfnamefont{J.~B.} \bibnamefont{Zuber}}
  \bibnamefont{and} \bibinfo{editor}{\bibfnamefont{R.}~\bibnamefont{Stora}}
  (\bibinfo{publisher}{North Holland, Amsterdam}, \bibinfo{year}{1984}), p.
  \bibinfo{pages}{473}.

\bibitem{tissier11b}
M. Tissier and G. Tarjus, preceding paper (2011).

\bibitem[{\citenamefont{Tarjus and Tissier}(2004)}]{tarjus04}
\bibinfo{author}{\bibfnamefont{G.}~\bibnamefont{Tarjus}} \bibnamefont{and}
  \bibinfo{author}{\bibfnamefont{M.}~\bibnamefont{Tissier}},
  \bibinfo{journal}{Phys. Rev. Lett} \textbf{\bibinfo{volume}{93}},
  \bibinfo{pages}{267008} (\bibinfo{year}{2004}).

\bibitem[{\citenamefont{Tissier and Tarjus}(2006)}]{tissier06}
\bibinfo{author}{\bibfnamefont{M.}~\bibnamefont{Tissier}} \bibnamefont{and}
 \bibinfo{author}{\bibfnamefont{G.}~\bibnamefont{Tarjus}},
  \bibinfo{journal}{Phys. Rev. Lett.} \textbf{\bibinfo{volume}{96}},
  \bibinfo{pages}{087202} (\bibinfo{year}{2006}).

\bibitem[{\citenamefont{Schwartz}(1985)}]{schwartz85b}
M. Schwartz, J. Phys. C {\bf 18}, 135 (1985).

\bibitem[{\citenamefont{Soffer}(1985)}]{soffer85}
M. Schwartz and A. Soffer, Phys. Rev. Lett. {\bf 55}, 2499 (1985); Phys. Rev. B {\bf 33}, 2059 (1986).

\bibitem[{\citenamefont{Schwartz}(1991)}]{schwartz91}
M. Schwartz, M. Gofman, and T. Natterman, Physica A {\bf 178}, 6 (1991).

\bibitem{tissier11}
M. Tissier and G. Tarjus, Phys. Rev. Lett. \textbf{107}, 041601 (2011).

\bibitem[{\citenamefont{Zinn-Justin}(1989)}]{zinnjustin89}
\bibinfo{author}{\bibfnamefont{J.}~\bibnamefont{Zinn-Justin}},
  \emph{\bibinfo{title}{Quantum Field Theory and Critical Phenomena}}
  (\bibinfo{publisher}{Oxford University Press}, \bibinfo{address}{New York},
  \bibinfo{year}{1989}), \bibinfo{edition}{3rd} ed.

\bibitem{wetterich93}
C. Wetterich, Physics Letters B \textbf{301}, 90 (1993).

\bibitem[{\citenamefont{Berges et~al.}(2002)\citenamefont{Berges, Tetradis, and
  Wetterich}}]{berges02}
\bibinfo{author}{\bibfnamefont{J.}~\bibnamefont{Berges}},
  \bibinfo{author}{\bibfnamefont{N.}~\bibnamefont{Tetradis}}, \bibnamefont{and}
  \bibinfo{author}{\bibfnamefont{C.}~\bibnamefont{Wetterich}},
  \bibinfo{journal}{Phys. Rep.} \textbf{\bibinfo{volume}{363}},
  \bibinfo{pages}{223} (\bibinfo{year}{2002}).

\bibitem[{\citenamefont{Cardy}(1983)}]{cardy83}
J. Cardy, Phys. Lett. B {\bf 125}, 470 (1983).

\bibitem{klein83}
A. Klein and J. F. Perez, Phys. Lett. B {\bf 125}, 473 (1983).

\bibitem[{\citenamefont{Klein et~al.}(1984)\citenamefont{Klein, Landau, and  Perez}}]{klein84}
\bibinfo{author}{\bibfnamefont{A.}~\bibnamefont{Klein}},
  \bibinfo{author}{\bibfnamefont{L.~J.} \bibnamefont{Landau}},
  \bibnamefont{and} \bibinfo{author}{\bibfnamefont{J.~F.} \bibnamefont{Perez}},
  \bibinfo{journal}{Commun. Math. Phys.} \textbf{\bibinfo{volume}{94}},
  \bibinfo{pages}{459} (\bibinfo{year}{1984}).
 
\bibitem[{\citenamefont{Footnote}(2)}]{footnote2}
We define ``supercusp'', ``cusp'', and ``subcusp'' in $\Gamma_{k2}^{(11)}( \phi_1,\phi_2)$ according to the order of the nonanalyticity in $|\phi_1-\phi_2|^{\alpha}$ when $\phi_2\rightarrow \phi_1$: $\alpha < 1$ corresponds to a supercusp, $1\leq \alpha <2$ to a ``cusp'', and $\alpha$ a noninteger $> 2$ to a subcusp. A ``linear cusp'' is associated with $\alpha = 1$. This can be generalized to higher-order cumulants of the renormalized random field.

\bibitem{larkin70}
A. I. Larkin, Sov. Phys. JETP \textbf{31}, 784 (1970).

\bibitem[{\citenamefont{Footnote}(3)}]{footnote3}
Note that there does not seem to be any alternative route, for instance by introducing new sources conjugate to operators that break SUSY  explicitly.We have not found any useful additional WT identities when the invariance under superrotations is broken. To check this, we have followed Zinn-Justin's procedure,\cite{zinnjustin89} which amounts to performing transformations on the action and adding sources to the new generated independent operators. However, introducing new operators to compensate for the terms breaking  SUSY breaks additional symmetries, including rotations in Euclidean subspace, and an increasing number of independent operators are generated at each new iteration of the procedure.

\bibitem[{\citenamefont{Footnote}(4)}]{footnote4}
If one also wishes to describe microscopic details associated with a specific definition of the bare action and keep a finite UV cutoff $\Lambda$, one has to generalize the function $s(q^2/k^2)$ to $s_{\Lambda}(q^2,k^2)=s(q^2/f_{\Lambda}(k^2))$, with $f_{\Lambda}(x) \simeq x$ when $\Lambda \rightarrow \infty$ and $f_{\Lambda}(x  \rightarrow \Lambda^2) \rightarrow \infty$; a  possible choice of $f_{\Lambda}(x)$ is $\Theta(\Lambda^2 - x) \Lambda^2 x/(\Lambda^2 - x)$ where $\Theta$ is the Heaviside step function. When $k \rightarrow \Lambda$ for $q^2$ fixed, the infrared cutoff function $\widehat{R}_{k}$ diverges as $f_{\Lambda}(k^2\rightarrow \Lambda^2)$ whereas $\widetilde{R}_{k}$ goes to $\Delta_B/2$, which fulfills the requirements described in paper III.\cite{tissier11b}

\bibitem[{\citenamefont{Footnote}(5)}]{footnote5}
The constraint of no explicit SUSY breaking and the associated WT identities are not easily implemented in other approximation schemes. For instance one could envisage to replace the derivative expansion by a more powerful scheme to describe the full momentum dependence of the proper vertices, as proposed in Ref.~[\onlinecite{BMW}]. However, it is not obvious then to combine this approximation with the expansion in cumulants.

\bibitem{BMW}
J.-P. Blaizot, R. Mendez-Galain, and N. Wschebor, Phys. Lett. B {\bf 632}, 571 (2006); Phys. Rev. E {\bf 74}, 051116 (2006); Phys. Rev. E {\bf 74}, 051117 (2006).

\bibitem{footnote51}
Note that if the bare random field is not Gaussian distributed, one cannot bluntly neglect all higher-order cumulants. One should then check that, as expected, the non-Gaussian behavior that is present at the bare level does not change the long-distance physics. To assess this point, it is necessary to include in the description at least the third cumulant and then consider the next order of the approximation scheme.

\bibitem{tetradis94}
N. Tetradis and C. Wetterich, Nucl. Phys. B \textbf{422}, 541 (1994).

\bibitem{delamotte03}
B. Delamotte, D. Mouhanna, and M. Tissier, Phys. Rev. B {\bf 69}, 134422 (2003).

\bibitem{ballhausen04}
H. Ballhausen, J. Berges, and C. Wetterich, Phys. Lett. B \textbf{582}, 144 (2004).

\bibitem{tissier06b}
M. Tissier and G. Tarjus, Phys. Rev. B \textbf{74}, 214419 (2006).

\bibitem{litim00}
D. F. Litim, Phys. Lett. B \textbf{486}, 92 (2000); Int. J. Mod. Phys. A \textbf{16}, 2081 (2001); Nucl. Phys. B \textbf{631}, 128 (2002).

\bibitem{canet03}
L. Canet, B. Delamotte, D. Mouhanna, and J. Vidal, Phys. Rev. D {\bf 67}, 065004 (2003).

\bibitem{pawlowski07}
J. M. Pawlowski, Ann. Phys. \textbf{322}, 2831 (2007).

\bibitem[{\citenamefont{Footnote}(6)}]{footnote6}
When $d>d_{DR}$, one anticipates a subcusp in $\delta_*(\varphi+y,\varphi-y)$ as $y \rightarrow 0$, with no implication for the dimensional reduction. Numerically however, it is very hard to find evidence for such subcusps as this would involve computing high orders in the derivatives with respect to the fields: for instance, one should be able to follow the fourth derivative of $\delta_k$ with respect to $y$ and check for its divergence at the fixed point above $d_{DR}$.

\bibitem{middleton-fisher02}
A. A. Middleton and D. S. Fisher, Phys. Rev. B \textbf{65}, 134411 (2002).

\bibitem{hartmann02}
A. K. Hartmann, Phys. Rev. B \textbf{65}, 174427 (2002).

\bibitem{middleton02}
A. A. Middleton, arXiv:cond-mat/0208182 (2002).

\bibitem{hartmann99}
A. K. Hartmann and U. Nowak, Eur. Phys. J. B \textbf{7}, 105 (1999).

\bibitem{ogielski86} 
A. T. Ogielski, Phys. Rev. Lett. \textbf{57}, 1251 (1986).

\bibitem{rieger93} 
H. Rieger and A. P. Young, J. Phys. A \textbf{26}, 5279 (1993).

\bibitem{rieger95} 
H. Rieger, Phys. Rev. B \textbf{52}, 5659 (1995).

\bibitem{fytas11}
N. G. Fytas and A. Malakis, Eur. Phys. J. B \textbf{79}, 13 (2011).

\bibitem{dukovski03} 
I. Dukovski and J. Machta, Phys. Rev. B \textbf{67}, 014413 (2003).

\bibitem{angles97} 
J.-C. Angles d'Auriac and N. Sourlas, Europhys. Lett. \textbf{39}, 473 (1997).

\bibitem{wu06} 
Y. Wu and J. Machta, Phys. Rev. B \textbf{74}, 064418 (2006).

\bibitem{hartmann01} 
A. K. Hartmann and A. P. Young, Phys. Rev. B \textbf{64}, 214419 (2001).

\bibitem{nowak98} 
U. Nowak, K. D. Usadel, and J. Esser, Physica A \textbf{250}, 1 (1998).

\bibitem{malakis06} 
A. Malakis and N. G. Fytas, Phys. Rev. E \textbf{73}, 016109 (2006).

\bibitem{dayan93}
I. Dayan, M. Schwartz, and A. P. Young, J. Phys. A \textbf{26}, 3093 (1993).

\bibitem{newman93}
M. E. J. Newman, B. W. Roberts, G. T. Barkema, and J. P. Sethna, Phys. Rev. B \textbf{48}, 16533 (1993).

\bibitem{cao93}
M. S. Cao and J. Machta, Phys. Rev. B \textbf{48}, 3177 (1993).

\bibitem{falikov95}
A. Falikov, A. N. Berker, and S. R. McKay, Phys. Rev. B \textbf{51}, 8266 (1995).

\bibitem{efrat03}
A. Efrat and M. Schwartz, Phys. Rev. E \textbf{68}, 026114 (2003).

\bibitem{gofman93}
M. Gofman, J. Adler, A. Aharony, A. B. Harris, and M. Schwartz, Phys. Rev. Lett. \textbf{71}, 1569 (1993).

\bibitem{bray85}
A. J. Bray and M. A. Moore, J. Phys. C \textbf{18}, L927 (1985).

\bibitem[{\citenamefont{Footnote}(7)}]{footnote7}
One can also check that the ``magnetic-field eigenvalue'' $y_m$  corresponding to the relevant direction associated with  odd eigenfunctions in the field arguments is analytically obtained in any dimension as $y_m=(1/2)(d-4+\bar \eta)$, which as expected is the dimension of the fields; the corresponding eigenfunctions $u_m(\varphi)$, $z_m(\varphi)$, and $v_m(\varphi_1,\varphi_2)$ are expressed in terms of the fixed-point functions, namely: $u_m(\varphi)=u'_*(\varphi)$, $z_m(\varphi)=z'_*(\varphi)$, and $v_m(\varphi_1,\varphi_2)=(\partial_{\varphi_1}+\partial_{\varphi_2}) v_*(\varphi_1,\varphi_2)$.

\bibitem{bray84}
A. J. Bray and M. A. Moore, J. Phys. C \textbf{17}, L463 (1984).

\bibitem{fisher-huse88}
D. S. Fisher and D. A. Huse, Phys. Rev. B \textbf{38}, 373 (1988).

\bibitem{parisi81}
G. Parisi and N. Sourlas, Phys. Rev. Lett. \textbf{46}, 871 (1981).

\bibitem{imbrie03}
D. C. Brydges and J. Z. Imbrie, J. Stat. Phys. \textbf{110}, 503 (2003); Ann. Math. \textbf{158}, 1019 (2003).

\bibitem{sethna05}
J. P. Sethna, K. A. Dahmen and O. Perkovic, in \textit{The Science of Hysteresis}, edited by G. Bertotti and I. Mayergoyz (Elsevier, Amsterdam, 2005), p. 107.

\bibitem{vives04}
F. J. Perez-Reche and E. Vives, Phys. Rev. B \textbf{70}, 214422 (2004).

\bibitem{liu09}
Y. Liu and K. A. Dahmen, Phys. Rev. E \textbf{79}, 061124 (2009).

\bibitem{gribov78}
V. N. Gribov, Nucl. Phys. B \textbf{139}, 1 (1978).

\bibitem{esposito04}
G. Esposito, D. N. Pelliccia, and F. Zaccaria, Int. J. Geom. Meth. Mod. Phys. \textbf{1}, 423 (2004).

\bibitem{Gribov77}
  V.~N.~Gribov,
  Nucl.\ Phys.\  B {\bf 139} (1978) 1.


\bibitem{Zwanziger89}
D.~Zwanziger, Nucl.\ Phys.\  B {\bf 323}, 513 (1989);  Nucl.\ Phys.\  B {\bf 399}, 477 (1993).


\end{thebibliography}
\end{document}